\documentstyle[psfig,amstex]{mn}
\def\la{Ly$\alpha$}
\def\etal{et~al.}
\def\spose#1{\hbox to 0pt{#1\hss}}
\def\lta{\mathrel{\spose{\lower 3pt\hbox{$\mathchar"218$}}
     \raise 2.0pt\hbox{$\mathchar"13C$}}}
\def\gta{\mathrel{\spose{\lower 3pt\hbox{$\mathchar"218$}}
     \raise 2.0pt\hbox{$\mathchar"13E$}}}
\def\Ha{H$\alpha$}
\def\Hb{H$\beta$}

\title[Radio galaxies at $z \sim 1$]{HST, radio and infrared observations
of 28 3CR radio galaxies at redshift ${\mathbf z \sim 1}$ --- II. Old
stellar populations in central cluster galaxies}

\author[P.~N.~Best \etal]{P.~N.~Best$^1$, M.~S.~Longair$^2$ and
H.~J.~A.~R\"ottgering$^1$ \\
$^1$ Sterrewacht Leiden, Huygens Laboratory, P.O. Box 9513,
2300 RA Leiden, The Netherlands \\ 
$^2$ Cavendish Laboratory, Madingley Road,
Cambridge, CB3 0HE, England}

\pagerange{\pageref{firstpage}--\pageref{lastpage}}
\pubyear{1995}

\begin{document}
 
\label{firstpage}
 
\maketitle
 
\begin{abstract}
\smallskip

\noindent Hubble Space Telescope images of 3CR radio galaxies at redshifts
$0.6 < z < 1.8$ have shown a remarkable variety of structures, generally
aligned along the radio axis, indicating that the radio source strongly
influences the optical appearance of these galaxies. In this paper we
investigate the host galaxies underlying this aligned emission, combining
the HST data with ground--based infrared images. 
\smallskip

\noindent An investigation of the spectral energy distributions of the
galaxies shows that the contribution of the aligned blue component to the
K--band light is generally small ($\sim 10$\%). The radial intensity
profiles of the galaxies are well matched at radii $\lta 35$\,kpc by de
Vaucouleurs' law, demonstrating that the K--band light is dominated by that
of an elliptical galaxy. There is no evidence for a nuclear point
source, in addition to the de Vaucouleurs profile, with a contribution
$\gta 15\%$ of the total K--band flux density, except in two cases, 3C22
and 3C41. We conclude that the K--band emission of the distant 3CR
galaxies is dominated by starlight. The magnitudes, colours and location
of the distant 3CR galaxies on the projected fundamental plane indicate
that their stellar populations formed at high redshift and have since been
evolving passively.
\smallskip

\noindent Large characteristic radii are derived for the 3CR galaxies,
indicating that they must be highly evolved dynamically, even at a
redshift of one.  At radii larger than $\sim 35$\,kpc, a combined galaxy
profile clearly shows an excess of emission as compared with de
Vaucouleurs' law, indicating that at least some of the galaxies possess
cD--type halos. This supports other independent evidence for the
hypothesis that the distant 3CR galaxies lie in moderately rich
(proto--)clusters.  Since the nearby FR\,II galaxies in the 3CR catalogue
lie in more diffuse environments and do not possess cD halos, the galactic
environments of the 3CR galaxies must change with redshift. The K$-z$
relation of the 3CR galaxies cannot, therefore, be interpreted using a
standard `closed--box, passively evolving stellar population' model,
whereby the galaxies that host distant 3CR sources will evolve into the
galaxies that host nearby 3CR FR\,II sources.
\smallskip

\noindent At redshifts $z \sim 1$, the absolute K--magnitudes of the
stellar populations of the 3CR galaxies are brighter than those of the
lower radio power 6C galaxies, indicating that the 3CR galaxies contain a
greater mass of stars; this is consistent with them lying towards the
centres of clusters. Powerful high redshift radio galaxies possess radio
beams whose kinetic power is close to the Eddington limiting luminosity of
a central supermassive black hole. Since the mass of the black hole is
likely to scale in proportion to the mass of the host galaxy, the 3CR
galaxies will contain more massive central engines than the 6C galaxies,
which accounts for their more powerful radio emission.  At redshifts $z
\lta 0.6$, the beam power of the radio sources is limited by the
availability of fuel for the central engine rather than the black hole
mass, and so no correlation is expected between the radio power and the
mass of the host galaxy.
\end{abstract}
\hfill

{\it Accepted for publication in Monthly Notices}
\clearpage


\section{Introduction}
\label{intro}

The 3CR sample of radio sources defined by Laing~\etal\ \shortcite{lai83}
consists of the brightest extragalactic radio sources in the northern sky,
selected at 178\,MHz, and contains radio galaxies and radio quasars with
redshifts up to $z \sim 2$. The host galaxies of the low redshift sources
in the sample are identified with giant elliptical galaxies containing old
stellar populations and, if the host galaxies of the high redshift sources
are also giant ellipticals, their stellar populations can be used to study
the evolution of this class of galaxy with cosmic epoch and consequently
to constrain models of galaxy formation and evolution.

Lilly and Longair \shortcite{lil82,lil84a} obtained infrared K--magnitudes
for an almost complete sample of 83 3CR galaxies with redshifts $0 < z <
1.6$, and constructed the K--magnitude {\it vs} redshift relation for
these objects. The resulting relation showed remarkably little scatter and
was interpreted as indicating that the 3CR host galaxies at redshift $z
\sim 1$ are indeed giant elliptical galaxies.  Lilly \& Longair
\shortcite{lil84a} also showed that, unless the deceleration parameter
were as large as $q_0 \sim 3.5$, the shape of the K$-z$ relation would not
be consistent with non--evolving stellar populations, but that at least
passive evolution is required. The K$-z$ relation suggests that the host
galaxies must have formed at large redshift, $z_{\rm f}\gta 3$; this age
is consistent with the red colours of the infrared emission of some of
these galaxies \cite{lil89}. Support for such an early formation epoch
comes from a very deep spectrum of 3C65, at redshift $z = 1.176$, which is
well--matched using a stellar population of at least 4~Gyr in age
\cite{sto95}, and also from the recent results of Dunlop \etal\
\shortcite{dun96}. 

In 1987, McCarthy \etal\ and Chambers \etal\ discovered that the optical
emission of these powerful radio galaxies tends to be aligned along the
axis of the radio source.\nocite{cha87,mcc87} Many models have been
proposed to account for this alignment effect, the most promising being
massive star formation induced by the passage of the radio jets (e.g. Rees
1989)\nocite{ree89}, scattering of light from an obscured active galactic
nucleus (AGN) by electrons or by dust (Cimatti \etal\ 1996,1997, Dey
\etal\ 1996, and references therein)\nocite{cim96,cim97,dey96} and nebular
continuum emission from warm line--emitting regions \cite{dic95}. This
complicates the use of these galaxies as cosmological probes.

A number of observations have also suggested that a proportion of the
K--band light of these galaxies may not be associated with starlight: (i)
the discovery of a weaker alignment effect at near--infrared wavelengths
(e.g. Eisenhardt and Chokshi 1990; Rigler \etal\ 1992; Dunlop \& Peacock
1993)\nocite{eis90,rig92,dun93}, which may be produced by the long
wavelength tail of a flat spectrum component responsible for the optical
alignment effect; (ii) the detection of broad \Ha\ emission from the
distant radio galaxy 3C22 at $z=0.938$ \cite{raw95,eco95} and a reported
detection of an unresolved central component in the K--band emission of
3C65 at $z = 1.176$ \cite{lac95}, suggesting that the infrared emission
may contain light directly from the central AGN; (iii) the observation
that infrared emission lines such as [SIII]~9532 can contribute a
significant percentage of the K--band flux \cite{raw91a}.

In addition, the K$-z$ relation of the 6C radio galaxies, which are
roughly a factor of five lower in radio luminosity than the 3CR sources,
tends to track that of the 3CR galaxies at low redshift, but at higher
redshift the 6C galaxies tend to be fainter, lying closer to the `no
evolution' line \cite{eal96,eal97}. This result is consistent with the
suggestion of Yates \etal\ \shortcite{yat86} that there is an intrinsic
correlation between the absolute infrared magnitude and the radio
luminosity of powerful radio galaxies at large redshifts. Eales \etal\
\shortcite{eal97} have argued that a proportion of the K--band emission of
the 3CR galaxies may be directly or indirectly associated with AGN
activity, the K$-z$ relation being caused, in part, by the correlation
of radio luminosity with redshift in the flux--limited 3CR sample.

In order to study the astrophysics of powerful distant radio galaxies, we
have selected an almost complete sample of 28 sources from the 3CR
catalogue, in the redshift range $0.6 < z < 1.8$, for observation by the
Hubble Space Telescope (HST). In a previous paper (Best \etal\ 1997;
hereafter Paper I)\nocite{bes97c} we presented the results of observations
of these sources at optical wavelengths using the HST, at 8.4\,GHz using
the Very Large Array radio interferometer (VLA), and at 1.2 and 2.2 $\mu$m
using the IRCAM3 array on the United Kingdom InfraRed Telescope
(UKIRT). The HST observations showed that the optical morphologies of the
majority of the radio galaxies bear little resemblance to giant elliptical
galaxies, being dominated instead by high surface brightness structures
elongated along the axes of the double radio sources (Paper I). In
interpreting the HST images, however, it should be appreciated that
standard elliptical galaxies possess very red colours and, at a redshift
of one, have low surface brightnesses at optical (rest--frame ultraviolet)
wavelengths (e.g. Giavalisco \etal\ 1996).\nocite{gia96a} Only regions of
enhanced ultraviolet emission appear prominently in the images.

We defer an analysis of the variety of optical structures and of the
alignment effect to the third paper in this series (Best \etal\, in prep.;
Paper III). Here, we restrict the investigation to the old stellar
populations of the host radio galaxies. The outline of this paper is as
follows. In Section~\ref{kzdata}, we describe the data reduction used in
the current analysis which has not already been discussed in Paper I. In
Section~\ref{iralign}, we address the problem of estimating the fraction
of the K--band light which is associated with a flat spectrum aligned
component.  In Section~\ref{vaucpro} we investigate the radial intensity
profiles of the galaxies. The relationship between the half--light radii
and the surface brightnesses of the galaxies is compared with the locus of
low redshift elliptical galaxies and brightest cluster galaxies in
Section~\ref{remu}.  In Section~\ref{seckzdiag} we derive an improved
K$-z$ relation for the narrow line radio galaxies.  The implications of
our results for the formation and evolution of massive galaxies and for
the cosmic evolution of the radio source population are discussed in
Section~\ref{discuss}, and we summarise our conclusions in
Section~\ref{concs}.

Throughout the paper, the GISSEL stellar synthesis codes of Bruzual and
Charlot \shortcite{bru93} are used. We adopt a Scalo
\shortcite{sca86} initial mass function with upper and lower mass
cut--offs of $65 M_{\odot}$ and $0.1 M_{\odot}$ respectively, and solar
metallicity.  Except where otherwise stated, a deceleration parameter $q_0 =
0.5$ and a Hubble constant $H_0 = 50$\,km\,s$^{-1}$\,Mpc$^{-1}$ are
adopted.

\section{The data}
\label{kzdata}

A variety of infrared properties of the galaxies were measured directly
from the infrared images and are presented in Table~\ref{irtab1} (see
also the data tables presented in Paper I). Table~\ref{irtab2} contains
properties of the galaxies which are derived in the text.

\begin{table}
\begin{tabular}{lclcccc}
Source &  $z$  & K--mag.& Error & Ellip. & D.P.A. &K$_{\rm corr}$\\
&&\multicolumn{2}{c}{[9$''$ diam. ap.]}&   &[deg]   &       \\
\hspace{2mm} (1)&(2)&\hspace{1mm} (3)&(4)&(5)&(6) & (7)   \\
3C13   & 1.351 & 17.52$^*$& 0.06&  0.12  &   12   & 17.43 \\
3C22   & 0.938 &  15.40 &  0.15 &  0.12  &   12   &  ---  \\
3C34   & 0.690 & 16.43$^*$& 0.05&  0.10  &   29   & 16.15 \\ 
3C41   & 0.795 &  15.68 &  0.04 &  0.21  &   54   &  ---  \\ 
3C49   & 0.621 &  16.15 &  0.15 &  0.20  &    8   & 16.22 \\
3C65   & 1.176 &  16.59 &  0.07 &  0.16  &   60   & 16.60 \\
3C68.2 & 1.575 &  17.49 &  0.12 &  0.28  &    4   & 17.66 \\
3C217  & 0.897 &  17.52 &  0.08 &  0.22  &   51   & 17.67 \\
3C226  & 0.820 &  16.52 &  0.05 &  0.12  &   76   & 16.73 \\
3C239  & 1.781 & 17.83$^*$& 0.06&  0.07  &   80   & 17.70 \\
3C241  & 1.617 &  17.45 &  0.08 &  0.18  &   26   & 17.51 \\
3C247  & 0.749 & 15.96$^*$& 0.10&  0.08  &   25   & 15.78 \\
3C252  & 1.105 &  17.32 &  0.07 &  0.22  &    1   & 17.69 \\
3C265  & 0.811 &  16.03 &  0.04 &  0.12  &   14   & 16.26 \\
3C266  & 1.272 & 17.88$^*$& 0.09&  0.10  &    3   & 17.95 \\
3C267  & 1.144 &  17.21 &  0.05 &  0.36  &   38   & 17.33 \\
3C277.2& 0.766 & 17.27$^*$& 0.05&  0.11  &   22   & 17.33 \\
3C280  & 0.996 & 16.99$^*$& 0.04&  0.10  &   68   & 16.85 \\
3C289  & 0.967 &  16.66 &  0.07 &  0.19  &   65   & 16.71 \\
3C324  & 1.207 &  16.99 &  0.06 &  0.24  &    8   & 17.11 \\
3C337  & 0.635 &  16.57 &  0.08 &  0.13  &    6   & 16.60 \\
3C340  & 0.775 &  16.91 &  0.08 &  0.14  &   16   & 16.98 \\
3C352  & 0.806 & 16.92$^*$& 0.05&  0.13  &   35   & 16.81 \\
3C356  & 1.079 & 17.50$^*$& 0.06&  0.06  &   15   & 17.41 \\  
3C368  & 1.132 & 17.03$^{**}$&0.15&0.32  &    3   & 17.34 \\ 
3C437  & 1.480 &  17.74 &  0.20 &  0.11  &   77   & 17.74 \\ 
3C441  & 0.708 & 16.42$^*$& 0.04&  0.02  &    0   & 16.26 \\  
3C470  & 1.653 & 18.02$^*$& 0.15&  0.32  &   83   & 17.77 \\ 
\end{tabular}
\caption[irtab1]{\label{irtab1} Measured properties of the radio galaxies
(see Paper I). Column 1 contains the 3C catalogue name of the source, with
its redshift in column 2. The K--magnitude measured from our UKIRT images,
is given in column 3, with the error in column 4. This was generally
measured through a 9 arcsec aperture (for those galaxies indicated by an
asterisk, a 5 arcsec diameter aperture was used to exclude nearby
companions; see Paper I), and has been corrected for galactic extinction
using the extinction maps of Burstein and Heiles (1982). For 3C368 (marked
$^{**}$) the K magnitude quoted is after subtraction of the M--star (see
Paper I). The ellipticity of the K--band image is given in column 5 and
the position angle offset from the radio axis in column 6. In column 7 we
tabulate the K$_{\rm corr}$ magnitudes obtained after scaling to a 9 arcsec
diameter aperture and correcting for any flat spectrum and point source
contributions (see text).}
\end{table}
\nocite{bur82a}

The photometric magnitudes of the galaxies were measured at a variety of
wavelengths (see Paper I), and were converted into broad--band flux
densities for the investigation of the spectral energy distributions
(SEDs) of the galaxies in Section~\ref{iralign}. Most of the optical flux
densities are influenced to some extent by emission lines which lie within
the passband of the filter.  In Paper I, we calculated the percentage of
the flux density that is associated with line emission for each of the HST
observations; these values are used to produce `true' broad--band
continuum flux densities. In addition, the contributions of the
[OIII]~4959,5007 and \Hb\ emission lines to the J--band flux densities of
those galaxies at redshifts $z \gta 1.1$ have been estimated from the
[OII]~3727 line flux, using line ratios taken from the combined radio
galaxy spectrum derived by McCarthy \shortcite{mcc93}, and have been
subtracted from the J--band magnitudes. No correction has been made for
line contamination in the K waveband, although this should produce little
error since strong lines such as \Ha\ do not enter the K--waveband until a
redshift of $z \sim 1.9$, which is greater than that of any of the sources
in the sample. The only emission line of any significance which may be
present in the K--band is the [SIII]~9532 line. Rawlings \etal\
\shortcite{raw91a} detected a significant ($\sim 15\%$) contribution to
the K--band flux density of 3C356 in this line, but in none of the other
galaxies they observed was its contribution greater than $\sim 5$ to 10\%,
and often the line was not detected. In addition, this line only falls in
the K--band for the eight galaxies with redshifts $1.05 < z < 1.55$, and
these galaxies do not have significantly brighter absolute K--magnitudes.

Another factor which influences the measured flux densities is reddening
by dust intrinsic to the galaxies. Since no accurate estimates have yet
been made of the mass, temperature and distribution of dust throughout
these galaxies, taking account of dust extinction is problematic.  It has
a negligible effect at the long wavelengths of the K--band emission, with
which most of this paper is concerned, but it will affect the SED fitting
in Section~\ref{fitspecs}.

The distribution of dust within elliptical galaxies can be divided into
two components. First, there is dust spread throughout the galaxies: this
affects the colours of the galaxies but has little effect upon their
morphologies. This dust is therefore impossible to detect in our
observations, and cannot be properly compensated for. We discuss this
extended dust briefly when modelling the SED in Section~\ref{fitspecs}.
The second dust component is that which lies in disks near the centre of
the galaxies; de Koff \etal\ \shortcite{dek96} showed that between 30 and
40\% of low redshift radio galaxies possess clear dust lanes. The most
prominent example of this phenomenon in the galaxies in our sample is the
large dust lane running across the central regions of 3C324. Dickinson
\etal\ \shortcite{dic96} showed that an extinction E(B$-$V)$\sim 0.3$ is
required in the central regions of this galaxy to obscure the galaxy
nucleus; at the wavelengths of the HST observations, this extinction
coefficient corresponds to removing about 70\% of the flux density from
the regions underlying the dust lane (e.g. de Vaucouleurs \& Buta,
1983).\nocite{vau83} For 3C324, this correction results in a 19\% increase
in the total flux density through the F702W filter, and 16\% through the
F791W filter.

Four other galaxies, 3C68.2, 3C252, 3C266 and 3C356, also show reasonably
convincing evidence for dust lanes, although not to the extent of
3C324\footnote{Note that there is marginal evidence for weaker dust lanes
in the central regions of other galaxies in the sample; however, these are
too weak for reliable estimates of their effect to be made and, in any
case, the increase in flux density would be $\lta 5\%$. De Koff \etal\
\shortcite{dek96} suggested that the frequency of dust lanes in the 3CR
radio galaxies with redshifts $0.1 < z < 0.5$ is invariant with redshift,
corresponding to about 30 to 40\% of all the sources. Our images would be
consistent with a similar percentage of dust lanes being present in the
high redshift 3CR galaxies.}. The much more compact dust lanes in these
sources, relative to that of 3C324, make accurate measurements of the
extinction coefficient impractical, and so we adopt a value of half of
that of 3C324. Repeating the above procedure for these galaxies provides
best estimates of the optical flux density corrections of, in order of
increasing wavelength for each source: 3C68.2 --- 7\%; 3C252 --- 7\%, 4\%;
3C266 --- 10\%, 5\%, 4\%; 3C356 --- 6\%, 4\%. These corrections have been
incorporated into the flux density measurements used in the next section.

\section{The Infrared Aligned Component}
\label{iralign}

\subsection{Fits to the spectral energy distributions}
\label{fitspecs}

The infrared observations presented in Paper I confirm the result that the
K--band emission of the powerful radio galaxies in our sample has a
tendency to be aligned along the radio axis \cite{eis90,rig92,dun93} but
it is a significantly weaker effect than in the optical waveband. This
aligned infrared emission may be the long wavelength tail of the flat
spectrum component responsible for the alignment effect in the optical
waveband, or may indicate that the old stellar population itself is
aligned with the radio source, as in the models of Eales \shortcite{eal92}
and West \shortcite{wes91,wes94}.

A relatively simple procedure has been adopted for estimating the
contribution of the optically aligned component to the infrared
alignment. For each galaxy an SED is produced which best matches the
observed broad--band flux densities, using the sum of two components. The
first component is an old stellar population, whose SED is derived using
the GISSEL stellar synthesis codes of Bruzual and Charlot
\shortcite{bru93}. We assume that the stars formed in a 1~Gyr burst
which occurred at redshift $z = 10$, and the stellar population then evolved
passively until the observed epoch; negligible changes result for the SED
at redshift $z \sim 1$ if a formation redshift of 5 or 50, or an
exponentially decreasing star formation rate are used instead. The second
component represents the aligned emission; the spectral shape of this
component depends upon the physical cause of the alignment effect.

For electron scattering, a power--law quasar spectrum is expected, $f_{\rm
\nu} \propto \nu^{\alpha}$, where the value of $\alpha$ is generally about
$-$0.7 to $-$1 for quasars (e.g. Baker \& Hunstead 1996).\nocite{bak96}
Due to the much lower scattering efficiency of electrons compared to dust
(e.g. di Serego Alighieri \etal\ 1994)\nocite{dis94a}, if electron
scattering is to dominate, then the mass of extended dust must be small,
and extinction by the extended dust will have little effect on the
spectral shape. If scattering by dust is entirely responsible for the
excess ultraviolet flux then, assuming the dust to be evenly distributed
throughout the galaxies, a few times $10^7 M_{\odot}$ of dust are
required, (e.g. Di Serego Alighieri 1994)\nocite{dis94a}. Light scattered
by this dust will have a spectral shape somewhat bluer than the incident
quasar power law \cite{cal95,cim96}, but dust outside the scattering
regions will redden the scattered emission, giving it a similar spectral
shape to electron scattering. A young starburst has a spectral shape which
is relatively flat in $f_{\rm \nu}$, whilst nebular continuum emission has
a much redder spectral shape below about rest--frame 3600A but provides
little K--band intensity. Each of these effects is likely to be present at
some level, and so the overall SED of the aligned component will be some
combination of these. For illustrative purposes, We will assume that one
third of the aligned emission at rest--frame 3600\AA\ is associated with
each of the three processes of scattering, star-formation, and nebular
continuum; this is well within the range of previous estimates
(e.g. Dickson \etal\ 1995, Cimatti \etal\ 1997, Tadhunter \etal\
1996).\nocite{dic95,cim97,tad96} This combination can be approximated from
optical wavelengths through to K--band by a single spectral shape, $f_{\rm
\nu} \propto \nu^{-0.4}$, which is in close agreement with the $f_{\rm
\nu} \propto \nu^{-0.2}$ rest--frame ultraviolet spectral shape determined
for the powerful radio galaxy 4C41.17 \cite{dey97}. This,  $f_{\rm
\nu} \propto \nu^{-0.4}$, is the spectral shape we adopt for the aligned
emission, and hereafter we refer to this component as the `flat spectrum
component'.

\begin{figure*}
\centerline{\psfig{figure=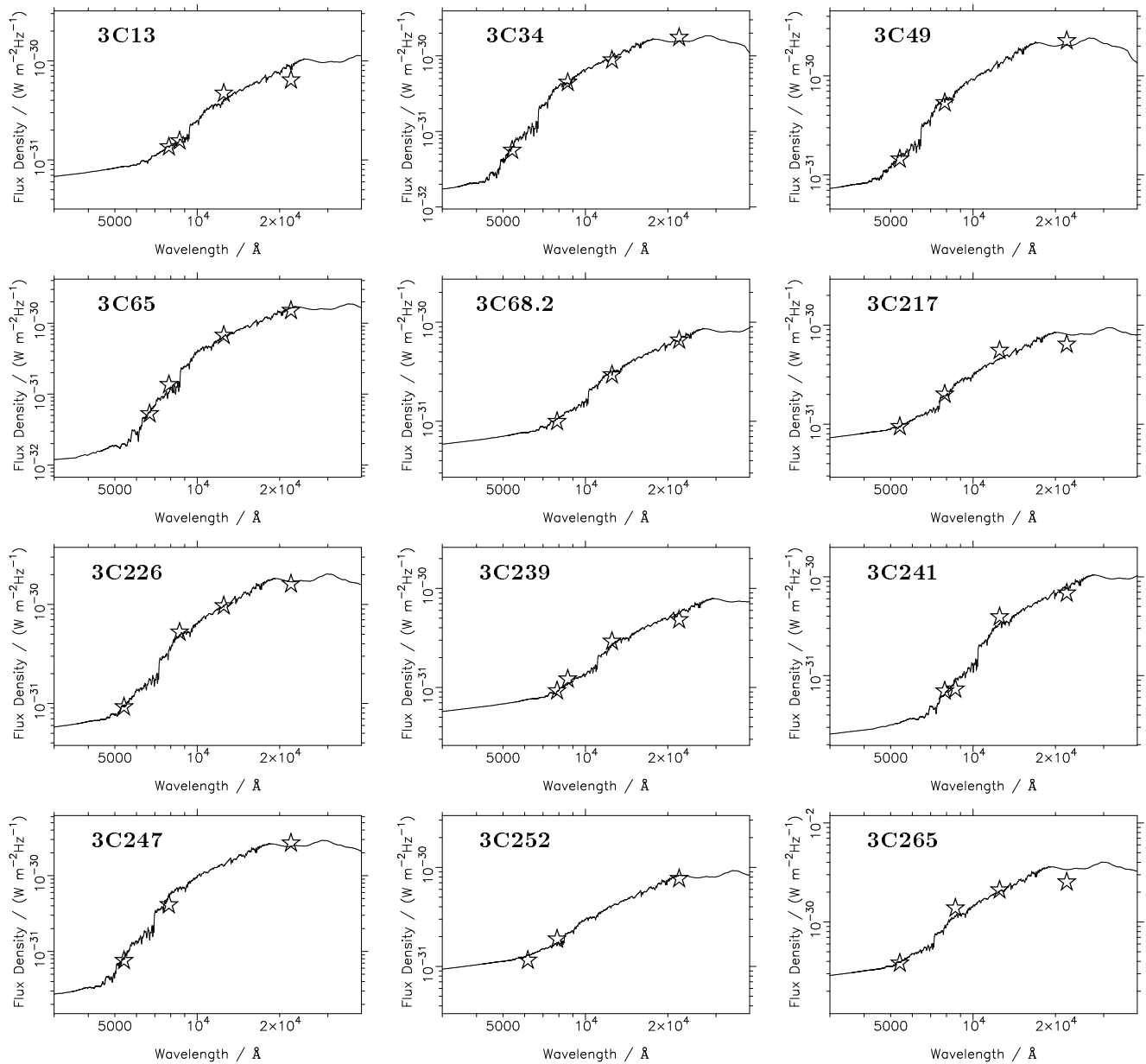,width=\textwidth,clip=}}
\caption{\label{specfits} SED fits to the broad band flux densities of the
3CR galaxies using an old stellar population and an aligned component (see
text and Table~\ref{irtab2} for details).}
\end{figure*}

\addtocounter{figure}{-1}
\begin{figure*}
\centerline{\psfig{figure=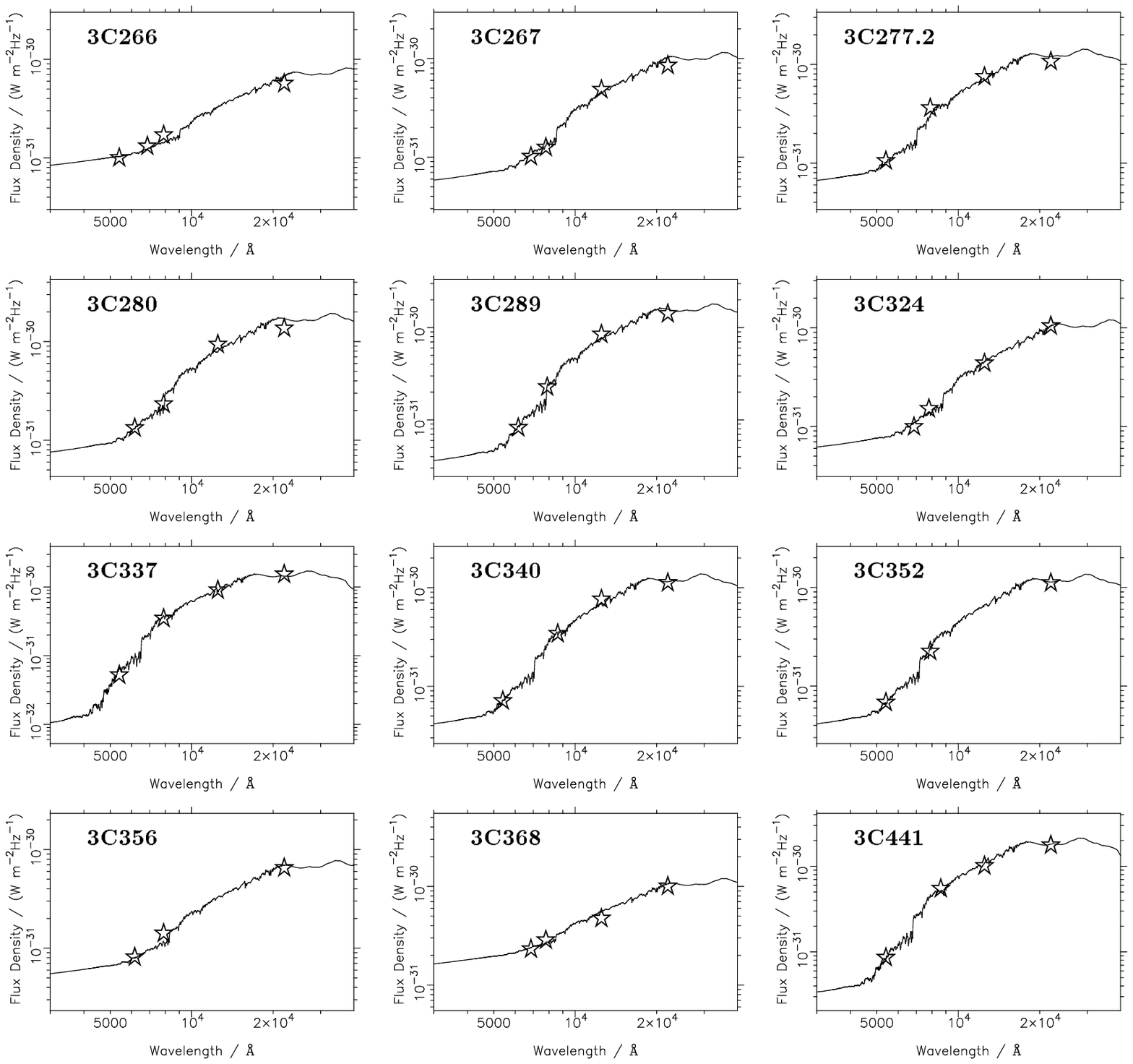,width=\textwidth,clip=}}
\caption{\label{specfitsb} cont.}
\end{figure*}

For each galaxy, the old stellar population and flat spectrum components
were weighted to produce the combined spectrum which best matched the
broad--band flux densities measured in the UKIRT and HST
observations. These fits are shown in Figure~\ref{specfits}. Such fits
were not made for 3C437 and 3C470, for which only two broad band flux
densities were available, nor for 3C22 and 3C41 which both possess
significant nuclear components in the infrared images (see
Section~\ref{vaucpro}).

The fits shown in Figure~\ref{specfits} are, in the majority of cases,
rather good, indicating that the simple two component model of the SED can
provide a good representation of the spectra of these galaxies. The mass
of stars within the galaxy and the flux density of the flat spectrum
component are listed in Table~\ref{irtab2}. Notice that the stellar masses
are typically a few times $10^{11} M_{\odot}$ which, given that the
mass--to--light ratios of such galaxies are usually well in excess of 1,
demonstrates that these galaxies are significantly more massive than
$M^*$.  The contributions to the K--band flux of the two components of the
fit can be compared to provide an estimate of the fraction of the emission
in this band which is associated with the aligned component. The values
derived from our best fitting models are tabulated in Table~\ref{irtab2}
and range from $\sim 1\%$ in the passive sources 3C65 and 3C337, up to
$\sim 31\%$ in the case of 3C368, with an mean value of about 10\%.

\begin{table*}
\begin{tabular}{lrrrrlccccc}
Source&Stellar\hspace{1.5mm} &Flat\,\,&Flat\hspace{2.7mm} &Align.\hspace{2.2mm} &\,$r_{\rm e}$& \%   & \%   & Point & $\mu_{\rm e}$&$\mu_{\rm e}$\\
      & Mass\hspace{2.8mm}   &Spec. &Spec.\hspace{2mm}  &Str.\hspace{3.5mm}   &             & Point& Point& Source&  K--band     & V--band     \\
      &                      &Flux\,&    (K--band)      &(K--band)            &             &Source&Source& Limits&              &(passive)    \\
      &[$10^{11}M_{\odot}$]  &[$\mu$Jy]&[\%]\hspace{3.5mm} &                  &\,[$''$] &[Measured]&[Corrected]&&\multicolumn{2}{c}{[Mags per sq. arcsec]}\\
\hspace{0.5mm} (1)&(2)\hspace{3.3mm} &(3)\hspace{1mm} &(4)\hspace{3.5mm} &(5)\hspace{4.5mm} &\,(6)   & (7) &  (8)    &  (9)    &  (10) & (11)  \\
3C13   & 5.5\hspace{3.5mm}  &  9.0\hspace{1mm} &  15\hspace{3.8mm}   &   0.09\hspace{3.5mm} &  1.25  &  0  &  0  &$  P<7  $& 19.91   &  22.31  \\
3C22   & ---\hspace{3.5mm}  &  ---\hspace{1mm} &  ---\hspace{3.8mm}  &   0.09\hspace{3.5mm} & 1.5    & 37  & 50  &$40<P<70$&--- &   ---   \\ 
3C34   & 4.2\hspace{3.5mm}  &  2.0\hspace{1mm} &   2\hspace{3.8mm}   &   0.04\hspace{3.5mm} &  3.9   &  5  &  5  &$  P<11 $& 21.10   &  23.96  \\ 
3C41   & ---\hspace{3.5mm}  &  ---\hspace{1mm} &  ---\hspace{3.8mm}  &$-0.04$\hspace{3.5mm} & 2.3    & 24  & 31  &$23<P<41$&--- &   ---   \\ 
3C49   & 4.5\hspace{3.5mm}  &  8.5\hspace{1mm} &   7\hspace{3.8mm}   &   0.16\hspace{3.5mm} &  1.05  &  0  &  0  &$  P<2  $& 18.32   &  21.24  \\
3C65   & 8.5\hspace{3.5mm}  &  1.5\hspace{1mm} &   1\hspace{3.8mm}   &$-0.05$\hspace{3.5mm} &  1.7   &  0  &  0  &$  P<8  $& 19.75   &  22.32  \\
3C68.2 & 5.5\hspace{3.5mm}  &  8.0\hspace{1mm} &  17\hspace{3.8mm}   &   0.26\hspace{3.5mm} &\, ---  & --- & --- &   ---   &  ---    &  ---    \\
3C217  & 2.6\hspace{3.5mm}  &  9.0\hspace{1mm} &  17\hspace{3.8mm}   &$-0.03$\hspace{3.5mm} &  2.3   &  0  &  0  &$  P<6  $& 21.47   &  24.24  \\
3C226  & 5.7\hspace{3.5mm}  &  7.0\hspace{1mm} &   6\hspace{3.8mm}   &$-0.08$\hspace{3.5mm} &  2.55  &  7  & 13  &$  P<21 $& 20.75   &  23.55  \\
3C239  & 5.8\hspace{3.5mm}  &  8.0\hspace{1mm} &  19\hspace{3.8mm}   &$-0.05$\hspace{3.5mm} &\, ---  & --- & --- &   ---   &  ---    &  ---    \\
3C241  & 7.7\hspace{3.5mm}  &  3.5\hspace{1mm} &   6\hspace{3.8mm}   &   0.08\hspace{3.5mm} &\, ---  & --- & --- &   ---   &  ---    &  ---    \\
3C247  & 7.5\hspace{3.5mm}  &  3.2\hspace{1mm} &   2\hspace{3.8mm}   &   0.04\hspace{3.5mm} &\, ---  & --- & --- &   ---   &  ---    &  ---    \\
3C252  & 3.1\hspace{3.5mm}  & 12.0\hspace{1mm} &  23\hspace{3.8mm}   &   0.22\hspace{3.5mm} &  1.4   &  8  & 17  &$  P<24 $& 20.41   &  23.06  \\
3C265  & 10.0\hspace{3.5mm} & 35.0\hspace{1mm} &  16\hspace{3.8mm}   &   0.08\hspace{3.5mm} &  1.9   &  4  &  8  &$  P<12 $& 19.65   &  22.45  \\
3C266  & 3.2\hspace{3.5mm}  & 11.0\hspace{1mm} &  24\hspace{3.8mm}   &   0.10\hspace{3.5mm} &  1.5   &  2  &  4  &$  P<13 $& 20.83   &  23.33  \\
3C267  & 4.5\hspace{3.5mm}  &  7.5\hspace{1mm} &  12\hspace{3.8mm}   &   0.06\hspace{3.5mm} &  1.7   &  0  &  0  &$  P<5  $& 20.48   &  23.10  \\
3C277.2& 3.5\hspace{3.5mm}  &  8.0\hspace{1mm} &  10\hspace{3.8mm}   &   0.06\hspace{3.5mm} &  2.0   &  8  & 15  &$  P<21 $& 20.83   &  23.64  \\
3C280  & 6.7\hspace{3.5mm}  &  9.5\hspace{1mm} &   9\hspace{3.8mm}   &$-0.05$\hspace{3.5mm} &  1.15  &  0  &  0  &$  P<8  $& 19.14   &  21.85  \\
3C289  & 6.5\hspace{3.5mm}  &  4.5\hspace{1mm} &   4\hspace{3.8mm}   &$-0.08$\hspace{3.5mm} &  2.45  &  1  &  1  &$  P<7  $& 20.65   &  23.38  \\
3C324  & 5.0\hspace{3.5mm}  &  8.0\hspace{1mm} &  12\hspace{3.8mm}   &   0.20\hspace{3.5mm} &  2.2   &  0  &  0  &$  P<4  $& 20.81   &  23.38  \\
3C337  & 3.5\hspace{3.5mm}  &  1.2\hspace{1mm} &   1\hspace{3.8mm}   &   0.11\hspace{3.5mm} &  1.15  &  2  &  2  &$  P<9  $& 18.90   &  21.81  \\
3C340  & 3.5\hspace{3.5mm}  &  5.0\hspace{1mm} &   7\hspace{3.8mm}   &   0.09\hspace{3.5mm} &  0.7   &  0  &  0  &$  P<4  $& 18.20   &  21.02  \\
3C352  & 3.7\hspace{3.5mm}  &  5.0\hspace{1mm} &   7\hspace{3.8mm}   &   0.03\hspace{3.5mm} &  2.0   &  4  &  7  &$  P<13 $& 20.31   &  23.11  \\
3C356  & 2.7\hspace{3.5mm}  &  7.0\hspace{1mm} &  16\hspace{3.8mm}   &   0.04\hspace{3.5mm} &  0.9   &  0  &  0  &$  P<14 $& 19.17   &  21.84  \\  
3C368  & 3.6\hspace{3.5mm}  & 21.0\hspace{1mm} &  31\hspace{3.8mm}   &   0.30\hspace{3.5mm} &\, ---  & --- & --- &  ---    &   ---   &  ---    \\ 
3C437  & 4.2$^*$\hspace{2.0mm} &3.5$^*$\hspace{-0.6mm} & ---\hspace{3.3mm} &$-0.08$\hspace{3.5mm} &\, --- & --- & --- & --- &  ---   &  ---    \\ 
3C441  & 5.0\hspace{3.5mm}  &  4.0\hspace{1mm} &   4\hspace{3.8mm}   &   0.02\hspace{3.5mm} &  1.65  &  0  &  0  &$  P<5  $& 19.35   &  22.24  \\  
3C470  & 4.0$^*$\hspace{2.0mm} &2.7$^*$\hspace{-0.6mm} & ---\hspace{3.3mm} &$-0.27$\hspace{3.5mm} &\, --- & --- & --- & --- &  ---   &  ---    \\
\end{tabular}
\caption[irtab2]{\label{irtab2} Properties of the radio galaxies derived
in the various sections of the paper. Column 1 contains the 3C catalogue
name of the source. Column 2 lists the stellar mass of the galaxy derived
from the fits in Section~\ref{fitspecs}, and column 3 the flux density at
rest--frame 3000\AA\ of the `flat spectrum' component derived from the
same fit. The values for 3C437 and 3C470 (marked with an asterisk) are
only estimates since only two broad band flux densities were
available. This analysis was not possible for 3C22 and 3C41 because of a
nuclear contribution to their flux densities. For the remaining 24
galaxies, the approximate percentage of the K--band light that is
associated with a flat spectrum aligned component is given in column
4. The K--band alignment strength of each galaxy, as defined in
Section~\ref{iralori}, is listed in column 5. Column 6 gives the
best--fitting de Vaucouleurs radius for each galaxy (see
Section~\ref{vaucpro}, where the errors involved are also discussed). Such
fits were not possible for 3C247 and 3C368, nor for the five highest
redshift sources (see text). The percentage of K--band light associated
with an unresolved nuclear emission source in the best--fitting model is
given in Column 7. The value after a correction factor has been applied
(see Section~\ref{vaucpro} for further details) is given in Column 8, and
the 90\% confidence limits for this value are given in column 9. The
measured mean surface brightness within the de Vaucouleurs radius in the
K--band is given in column 10, and the modelled V--band surface brightness
for a passive evolution model is listed in column 11 (see
Section~\ref{remu} for more details). This analysis was not carried out
for 3C22 and 3C41 because of the presence of a nuclear contribution to the
K--band flux.}
\end{table*}

\subsection{The origin of the aligned infrared emission}
\label{iralori}

\begin{figure}
\centerline{\psfig{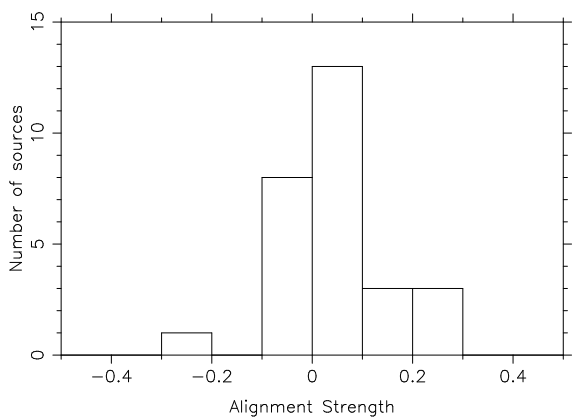}}
\caption{\label{kalign} A histogram of the K--band alignment
strengths. Values lying on a boundary are included in the lower bin.}
\end{figure}

To quantify the degree of alignment, we define the {\it alignment strength}
as:

\begin{displaymath}
a_{\rm s} = \epsilon \left (1 - \frac{\Delta\theta}{45}\right ),
\end{displaymath}

\noindent where $\Delta\theta$ is the difference in position angle between
the K--band emission and the radio axis, measured in degrees, and
$\epsilon$ is the ellipticity of the K--band image, defined as:

\begin{displaymath}
\label{epsilon}
\epsilon = \left[1 - \left({\overline{q^2}}/{\overline{p^2}}\right)\right]^{1/2}, 
\end{displaymath}

\noindent where $p$ and $q$ are dimensions along the major and minor axes
respectively of the infrared image of the radio galaxy. The alignment
strength therefore ranges from +1 to $-1$: those galaxies in which the
major axis of the infrared emission is oriented within 45$^{\circ}$ of the
radio axis have a positive alignment strength, whilst those which are
misaligned have a negative value; a value of zero corresponds either to a
galaxy whose major axis is misaligned by 45$^{\circ}$ from the radio axis
or to a galaxy which is circularly symmetric.  This is a better statistic
than considering $\Delta\theta$ alone since it results in a higher value
for a source which is highly elongated but misaligned by, say,
15$^{\circ}$ from the axis of the radio emission than for a source which
is virtually symmetric but whose major axis lies directly along the radio
axis. The values of the K--band alignment strengths are tabulated in
Table~\ref{irtab2} and are plotted as a histogram in
Figure~\ref{kalign}. There is a clear tendency for the galaxies to show
some alignment in the K waveband, with nearly 70\% possessing positive
alignment strengths, a result significant at greater than the 95\%
confidence level. The mean value of the alignment strength is 0.05.

\begin{figure}
\centerline{
\psfig{figure=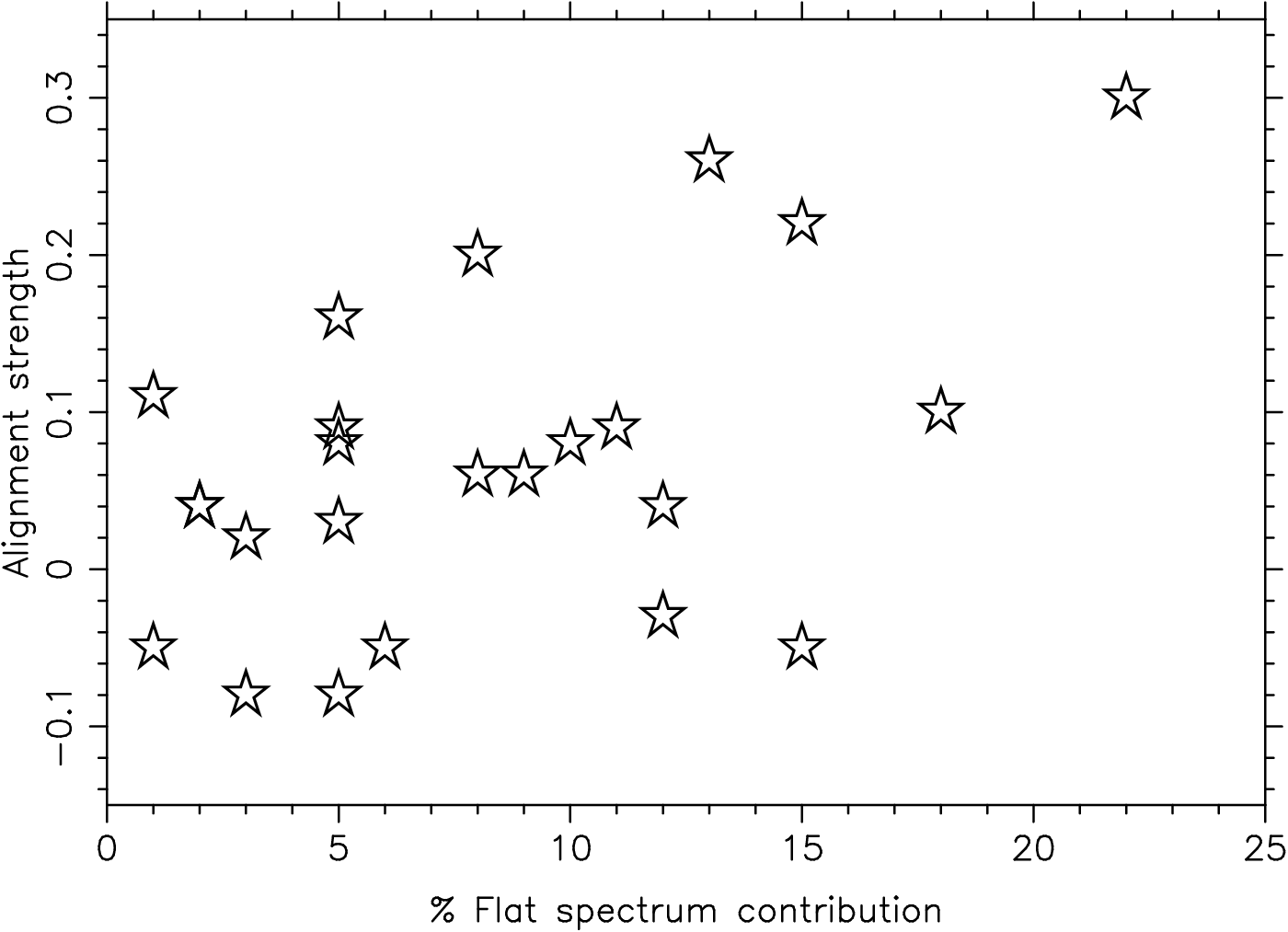,clip=,angle=-90,width=7.5cm}
}
\caption{\label{aliflat} A plot of the K--band alignment strength against
the percentage contribution at K of the flat spectrum component required
to match the SED of the galaxy.}
\end{figure}

It is interesting to compare the K--band alignment strength with the
percentage flat spectrum contribution in the K--band
(Figure~\ref{aliflat}). Of the five galaxies with the largest K--band flat
spectrum components, four are amongst the seven sources with high K--band
alignment strengths, $a_{\rm s} \ge 0.10$, including all three in the
highest bin in Figure~\ref{kalign} (see also Table~\ref{irtab2}). A
Spearmann rank test indicates that these parameters are correlated, at a
significance level of about 97.5\%.  If the four sources, 3C68.2, 3C252,
3C266, and 3C368, with particularly high values of both $a_{\rm s}$ and
the flat spectrum contribution are excluded, then for the remaining 24
galaxies the average alignment strength is $\left<a_{\rm s}\right> =
0.02$, different from the hypothesis of no alignment at only low
significance.

The correlation displayed in Figure~\ref{aliflat} indicates that the long
wavelength tail of the active blue emission responsible for the alignment
of the optical images is an important source of the K--band alignment
effect. Scatter in this relation will arise from the errors introduced in
determining the two parameters from the data and, more importantly, from
any intrinsic scatter in the alignment strength, such as that suggested by
the negative tail in Figure~\ref{kalign}. An estimate of the underlying
scatter can be gauged from the range of alignment strengths in
Figure~\ref{aliflat} of those galaxies with only small flat spectrum
components, suggesting a standard deviation of between 0.05 and 0.1. If
the parent galaxies are giant elliptical galaxies, this can naturally be
explained in terms of the radio axis being randomly orientated with
respect to the major axis of the galaxy.

\section{Radial Light Profiles}
\label{vaucpro}

\begin{figure*}
\centerline{
\psfig{figure=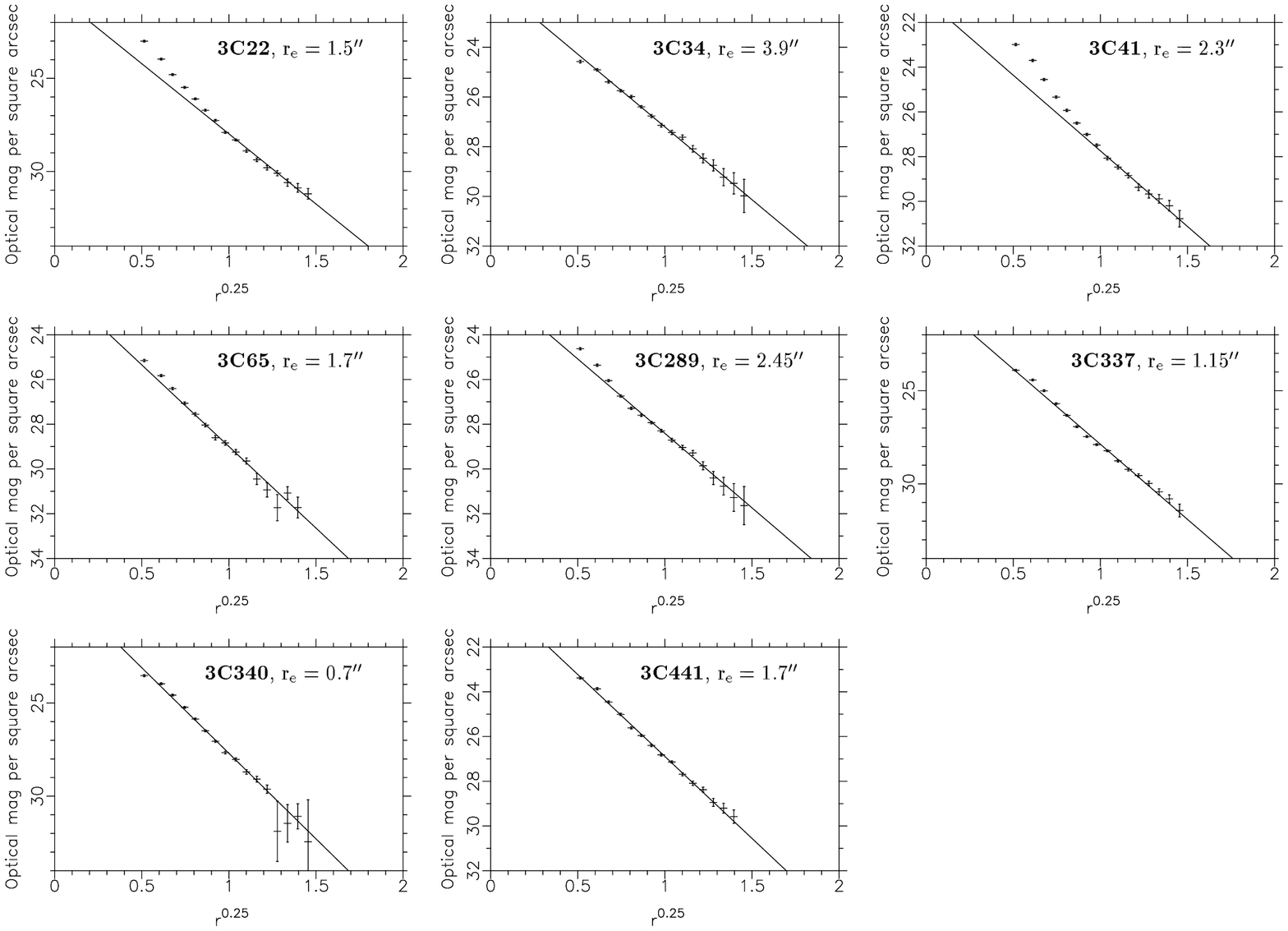,clip=,width=\textwidth,angle=-90}
} 
\caption{\label{hstvaucs} De Vaucouleurs fits to the radial intensity
profiles of the HST images of eight 3CR radio galaxies which do not show a
significant active ultraviolet component. The characteristic radius of
each, determined from the gradient of the best fitting straight line, is
given.}
\end{figure*}
\begin{figure*}
\centerline{
\psfig{figure=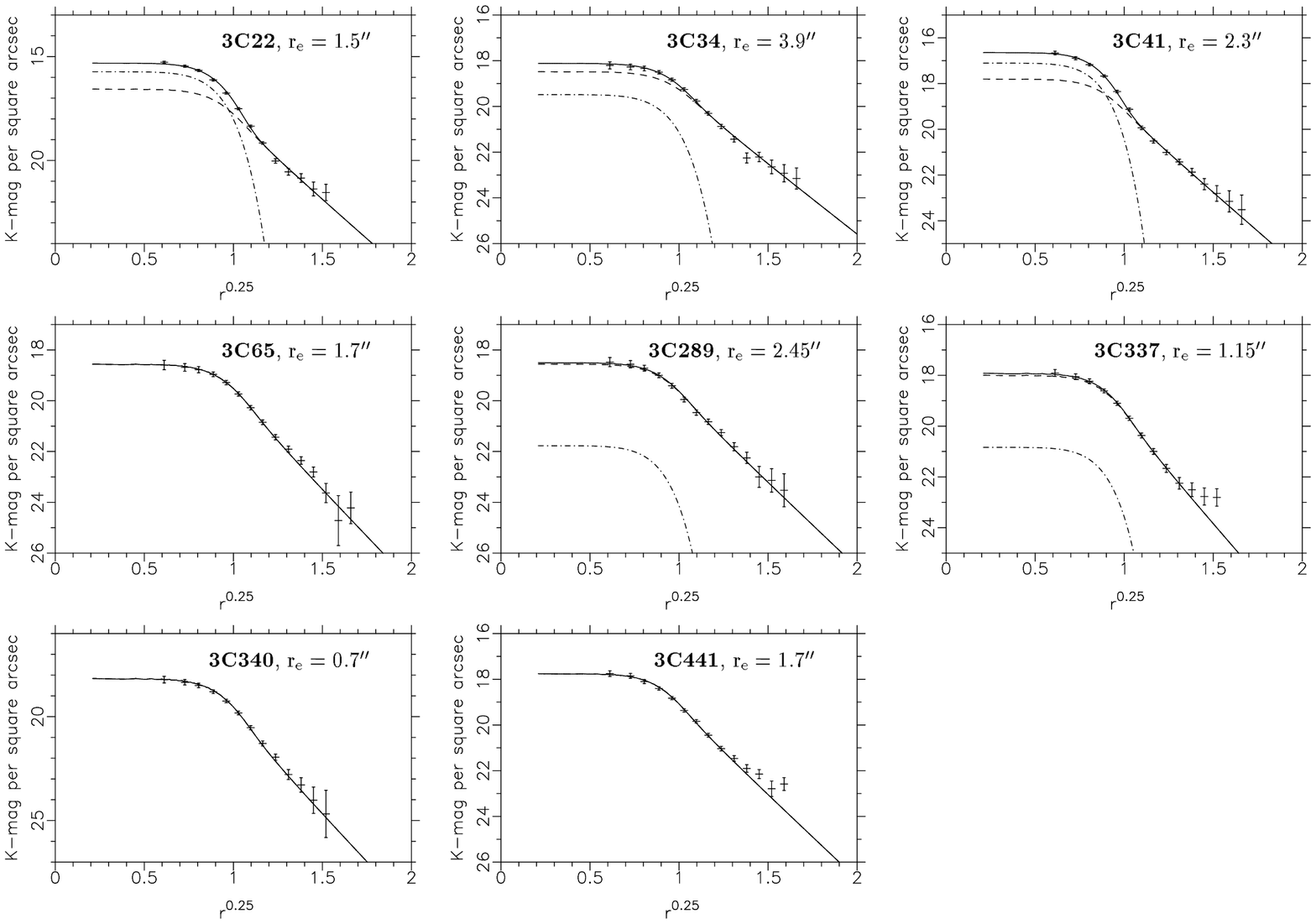,clip=,width=\textwidth,angle=-90}
}
\caption{\label{kvaucs} Fits to the radial intensity profiles of the
K--band images of the eight 3CR galaxies shown in Figure~\ref{hstvaucs},
using the sum of an unresolved point source (dash--dot line) and a de
Vaucouleurs profile with the characteristic radius determined from the HST
images (dashed line). For each of the profiles, the effect of seeing has
been taken into account. The sum of the two components is indicated by the
solid line. In many cases the best fit does not involve a point source
component, and so only the solid line is shown.}
\end{figure*}

Our observations can be used to test whether the K--band light
distributions of the galaxies have the profiles of elliptical galaxies, by
attempting to fit de Vaucouleurs' law, $I(r) \propto {\rm exp} \left
[-7.67 \left(r/r_{\rm e}\right )^{1/4}\right ]$, to their radial
surface brightness profiles. The HST images allow such plots to be made
without the need to account for the effects of seeing, and would allow an
excellent determination of the radial profiles to be made, were it not for
the fact that at optical wavelengths the emission from the majority of the
galaxies is dominated by the aligned blue component and so this analysis
cannot be carried out. Eight of the galaxies, however, do not show strong
ultraviolet emission. The HST images show that they possess almost
symmetrical morphologies and only weak aligned components are required to
fit their SEDs ($\lta 10$\% of the total flux density in the HST
image). The radial intensity profiles of the optical emission of these
galaxies were measured, nearby companions objects being removed and
replaced with the average background intensity at that distance from the
centre of the galaxy. No companion subtended an angle greater than 20
degrees about the centre of the galaxy, and so any errors in this
replacement will be small. The radial profiles are displayed in
Figure~\ref{hstvaucs}. They show that a de Vaucouleurs profile provides an
excellent fit to the observed radial surface brightness in six of the
eight cases, providing some of the strongest evidence to date that these
high redshift galaxies are indeed giant elliptical galaxies. The derived
values of $r_{\rm e}$ are listed in Table~\ref{irtab2}, and have a typical
accuracy of about 15\%.  The cases of 3C22 and 3C41, the two sources for
which a good de Vaucouleurs fit is not obtained at small radii, are
discussed below.

The characteristic radii, $r_{\rm e}$, obtained from these fits were then
used to investigate the infrared K--band radial intensity profiles. Where
companion galaxies subtended less than about 45 degrees with respect to
the centre of the galaxy, these were removed as in the HST images, and
replaced by the average of the background pixels at that point. In two
cases, 3C34 and 3C441, nearby companion galaxies subtended an angle $\gta
45^{\circ}$ about the centre of the 3CR galaxy, and this substitution
could lead to large errors in the radial profiles. In these cases, the
radial intensity profile out to a radius of about one arcsec was
determined using the normal method, but at larger radii only the 2 or 3
quadrants around the source which did not contain the companion were used,
and the emission from these was then scaled to derive the average
intensities. The two methods matched well at the boundary at 1 arcsec
radius.

The infrared radial intensity profiles of these eight galaxies were
modelled using a two--component fit, involving a de Vaucouleurs profile
with the characteristic radius determined from the fits to the HST images,
and a point source of emission; each profile was convolved with a Gaussian
profile matching the seeing of the infrared observations. The seeing,
which was typically between 1 and 1.2 arcsec, was measured using objects
on the infrared images of the radio galaxy which appeared unresolved in
the HST data. The fitting procedure involved minimising the sum of the
squares of the offsets of the combined model profile from the data points
within the inner 3 arcsecond radius, weighting each data point inversely
by its error. Tests showed that the procedure converged to the same minima
from a complete range of initial values.

The resulting fits are shown in Figure~\ref{kvaucs}: the dashed line shows
the de Vaucouleurs profile, the dotted line shows the radial profile of
the point source, and the solid line shows the total intensity profile. In
many cases, the point source component is zero, and so only the solid line
is plotted. In each case a good match is obtained. The percentage of the
total K--band emission of each source associated with a point source for
the best fitting combination is given in Table~\ref{irtab2}.

This analysis indicates the reason for the poorness of fit of the de
Vaucouleurs profiles to the optical radial intensity profiles of 3C22 and
3C41: the fits for these galaxies indicate the presence of an unresolved
nuclear component contributing 37\% and 24\% of the total K--band flux
respectively. It is interesting that these galaxies are the most luminous
in our sample in the K--band, lying furthest from the mean K$-z$ relation.
Rawlings \etal\ \shortcite{raw95} have already suggested that 3C22 is a
reddened quasar observed close to the radio galaxy\,/\,quasar divide,
based upon the compactness of the infrared emission, the detection of
broad \Ha\ emission in their spectrum and that of Economou \etal\
\shortcite{eco95}, and the very red colour of its spectral energy
distribution. This hypothesis is supported by the fact that this source
possesses a bright radio core and a strong one--sided jet, features which
are characteristic of quasars, but not generally of radio galaxies (see
Paper I for further discussion). No such claim has yet been made for 3C41,
although we note that it is one of the few sources in the sample to
possess a clear radio jet (Paper I).

\begin{figure}
\centerline{
\psfig{figure=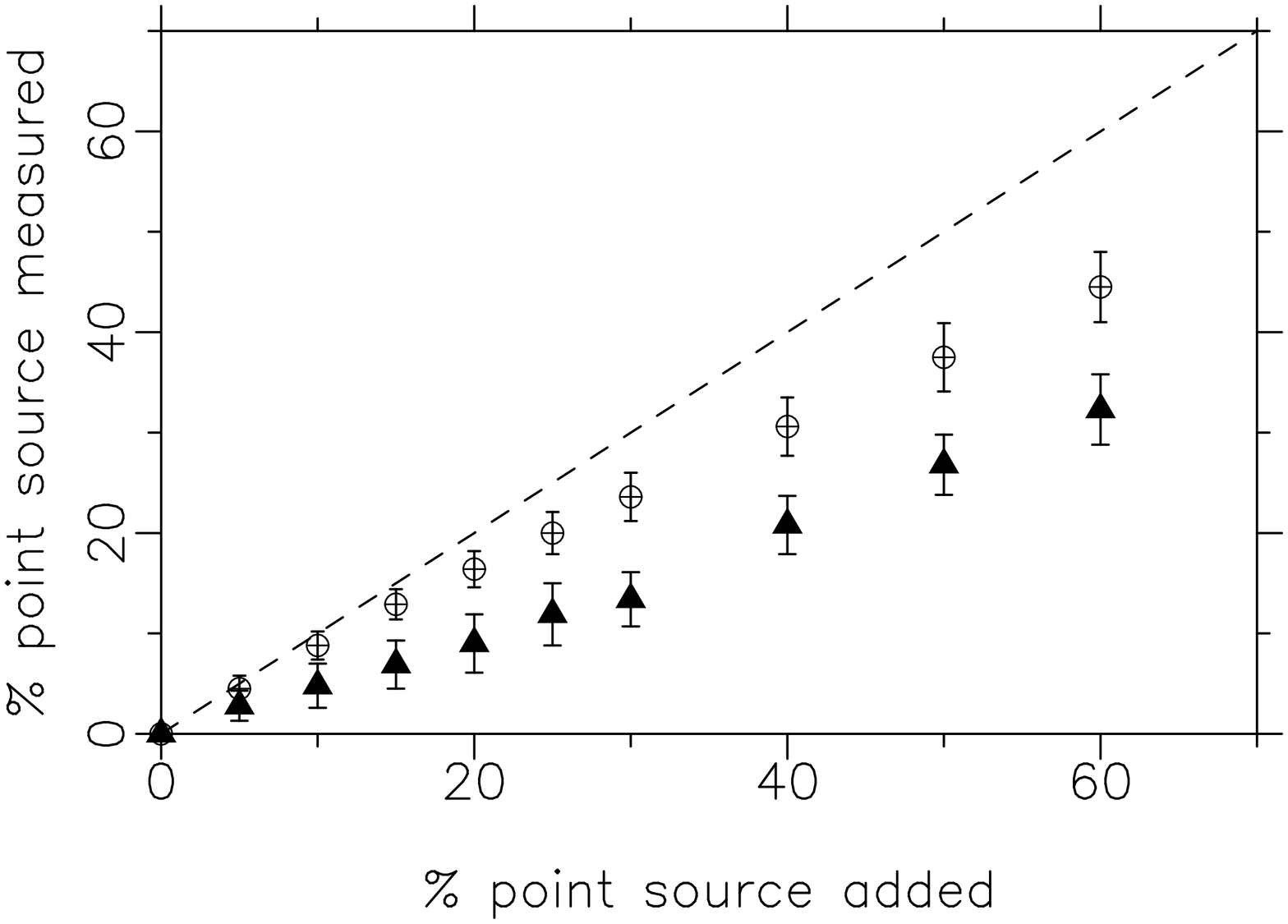,clip=,angle=-90,width=7.8cm}
}
\caption{\label{ptfig} The percentage of point source component determined
by the fitting routine for a given percentage of star added to the radio
galaxy. The crossed circles show the results when the characteristic
radius was known from the HST data, and the filled triangles show the
corresponding values for the galaxies for which the characteristic radius
was a free parameter in the fit.}
\end{figure}

Given the small size of these galaxies relative to the seeing profile, to
test the accuracy of the fitting procedure for determining the nuclear
point source contribution, we added the emission from a star, at various
percentages of the total flux density, to the infrared images of a number
of the radio galaxies. The resultant images were passed through our
fitting procedure to determine the percentage nuclear point source
recovered. The results are given in Table~\ref{pttab} and plotted in
Figure~\ref{ptfig} (crossed circles). For additions of a star of up to
about 15\% of the total flux density, the recovery of the point source
component is almost total, giving confidence to the result that there is
little or no requirement for a nuclear contribution in six of the galaxies
in Figure~\ref{kvaucs}. In particular, there is no evidence, either in the
de Vaucouleurs fit or in the SED, for the strong ($\sim 45\%$) nuclear
contribution to the emission of 3C65 suggested by Lacy \etal\
\shortcite{lac95}. For larger levels of star added, typically only about
75\% of the flux density is recovered as a point source. These results
suggest that the true point source contribution for 3C41 will be
approximately 30\%, and for 3C22 the best--fitting value will be about
50\%, with an upper limit of as much as about 70\% of the emission being
associated with an unresolved nuclear component. The `corrected' point
source contributions are given in Table~\ref{irtab2}, together with the
90\% confidence limits to this value as determined from the least--squares
fits.

\begin{table}
\begin{tabular}{cccc}
\% star added &\multicolumn{3}{c}{\% point source recovered} \\
              &Known $r_{\rm e}$  &~~~& Variable $r_{\rm e}$ \\
   5          &  4.5 $\pm$ 1.3    &&   2.8 $\pm$ 1.5         \\
  10          &  8.8 $\pm$ 1.4    &&   4.8 $\pm$ 2.2         \\
  15          & 12.9 $\pm$ 1.5    &&   6.9 $\pm$ 2.4         \\
  20          & 16.4 $\pm$ 1.8    &&   9.0 $\pm$ 2.9         \\
  25          & 20.0 $\pm$ 2.1    &&  11.9 $\pm$ 3.1         \\
  30          & 23.6 $\pm$ 2.4    &&  13.4 $\pm$ 2.7         \\
  40          & 30.6 $\pm$ 2.9    &&  20.8 $\pm$ 2.9         \\
  50          & 37.5 $\pm$ 3.4    &&  26.8 $\pm$ 3.0         \\
  60          & 44.5 $\pm$ 3.5    &&  32.3 $\pm$ 3.5         \\
\end{tabular}
\caption{\label{pttab} The percentage of point source component determined
by the fitting routine for a given percentage of star added to the radio
galaxy. This was carried out both for fitting where the characteristic
radius was known from the HST data, and when it was a free parameter in
the fit.}
\end{table}

Radial intensity profile fits can also be made for those galaxies which
possess enhanced ultraviolet emission in the HST images, since these too
are dominated by the underlying old stellar population at infrared
wavelengths. 3C368 and 3C247 were omitted from this analysis, the former
because of the presence of a foreground star within the galaxy envelope
which cannot be accurately removed (see Paper I), and the latter because
it is surrounded by nearby companions, too many to employ the quadrant
method discussed above. The 5 highest redshift galaxies in the sample were
also omitted due to low signal--to--noise ratios of the infrared emission
at large radii.

\begin{figure*}
\centerline{
\psfig{figure=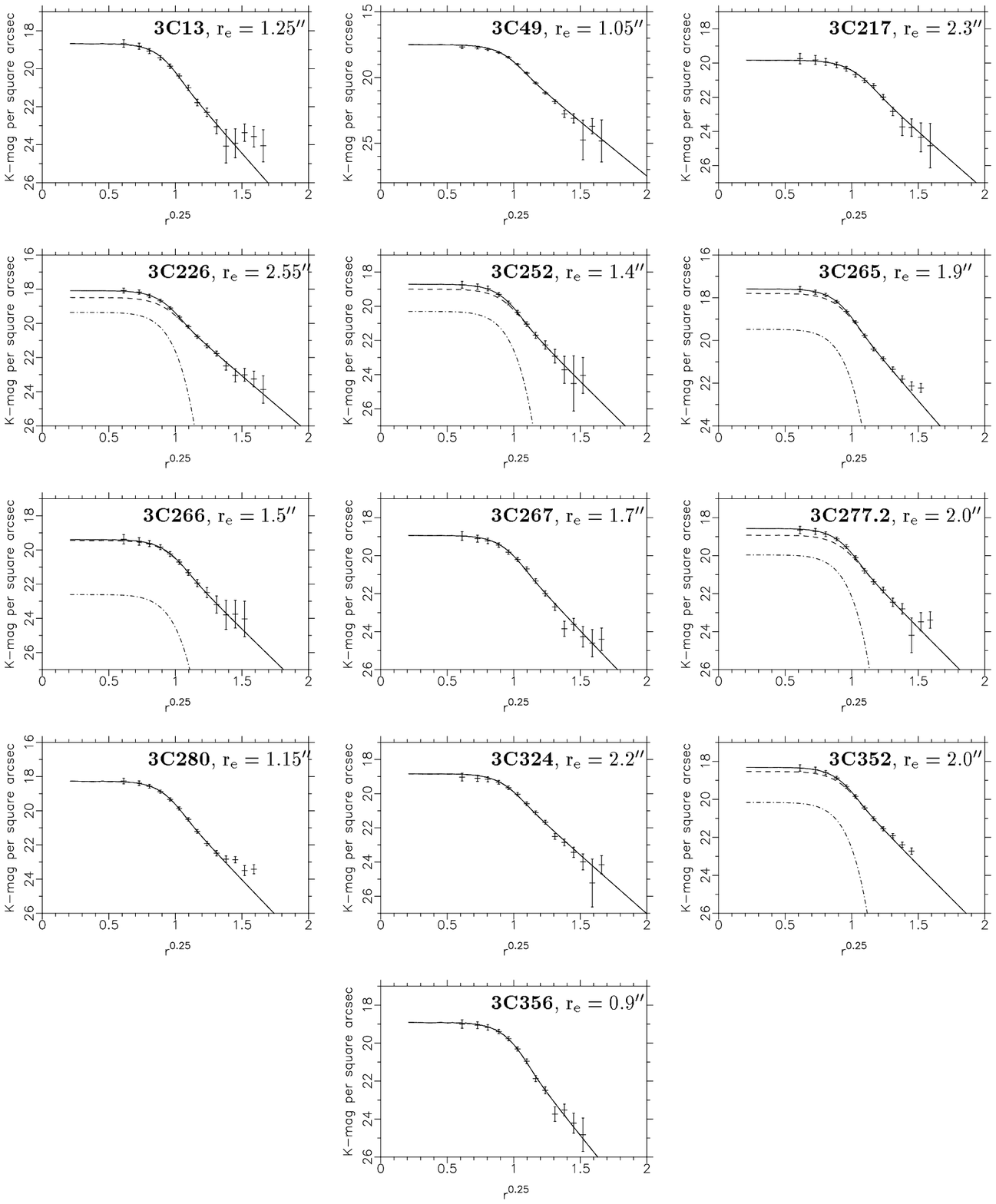,clip=,width=\textwidth}
}
\caption{\label{irvauc} Fits to the radial intensity profiles of the
K--band images of the 13 3CR galaxies for which de Vaucouleurs profiles
could not be determined from the HST observations. The characteristic
radii have been determined from the best fitting profiles. The same
notation is used as in Figure~\ref{kvaucs}.}
\end{figure*}

For the remaining 13 galaxies the infrared radial intensity profiles were
measured, and the best--fitting matches were derived from the sum of an
unresolved point source and a de Vaucouleurs profile. The characteristic
radius of the de Vaucouleurs profile was allowed to vary to obtain the
best match. The best--fitting models are compared with the observations in
Figure~\ref{irvauc}, with the same notation as in Figure~\ref{kvaucs}, and
the parameters of the models are given in Table~\ref{irtab2}. Although the
individual values of $r_{\rm e}$ can only be determined to a typical
accuracy of about 35\%, the galaxy light profiles can again be well fitted
using this model. 

The small size of the galaxies leads to some degeneracy between the
characteristic radius and the point source contribution. We have repeated
the test of adding a point source to the radio galaxies, allowing the
characteristic radius $r_{\rm e}$ to vary in the fit. The results are
given in Table~\ref{pttab} and shown in Figure~\ref{ptfig}. The
contribution of point source measured was lower than was added throughout,
typically by about a factor of two. What is happening is that the fitting
routine is estimating a slightly smaller value for $r_{\rm e}$, and
attributing the remainder of the point source flux to the galaxy
profile. The error on the points in Figure~\ref{ptfig} represents the
scatter in the results for addition to different galaxies, and arises
predominantly from the range of characteristic radii in the sample: the
recovery of the point source is least effective for those sources with the
smallest characteristic radii. However, even for small additions of a
stellar component, some point source contribution was generally recovered,
and therefore the very low point source percentages required for the
galaxies in Figure~\ref{irvauc} (mean value $\lta 4\%$) give confidence
that any nuclear contribution must be small. The `corrected' contributions
of the unresolved emission to the K--band flux density are listed in
Table~\ref{irtab2} together with the 90\% confidence limits on these
values.  Each of these galaxies is consistent with no point source
component being present.

\begin{figure}
\centerline{
\psfig{figure=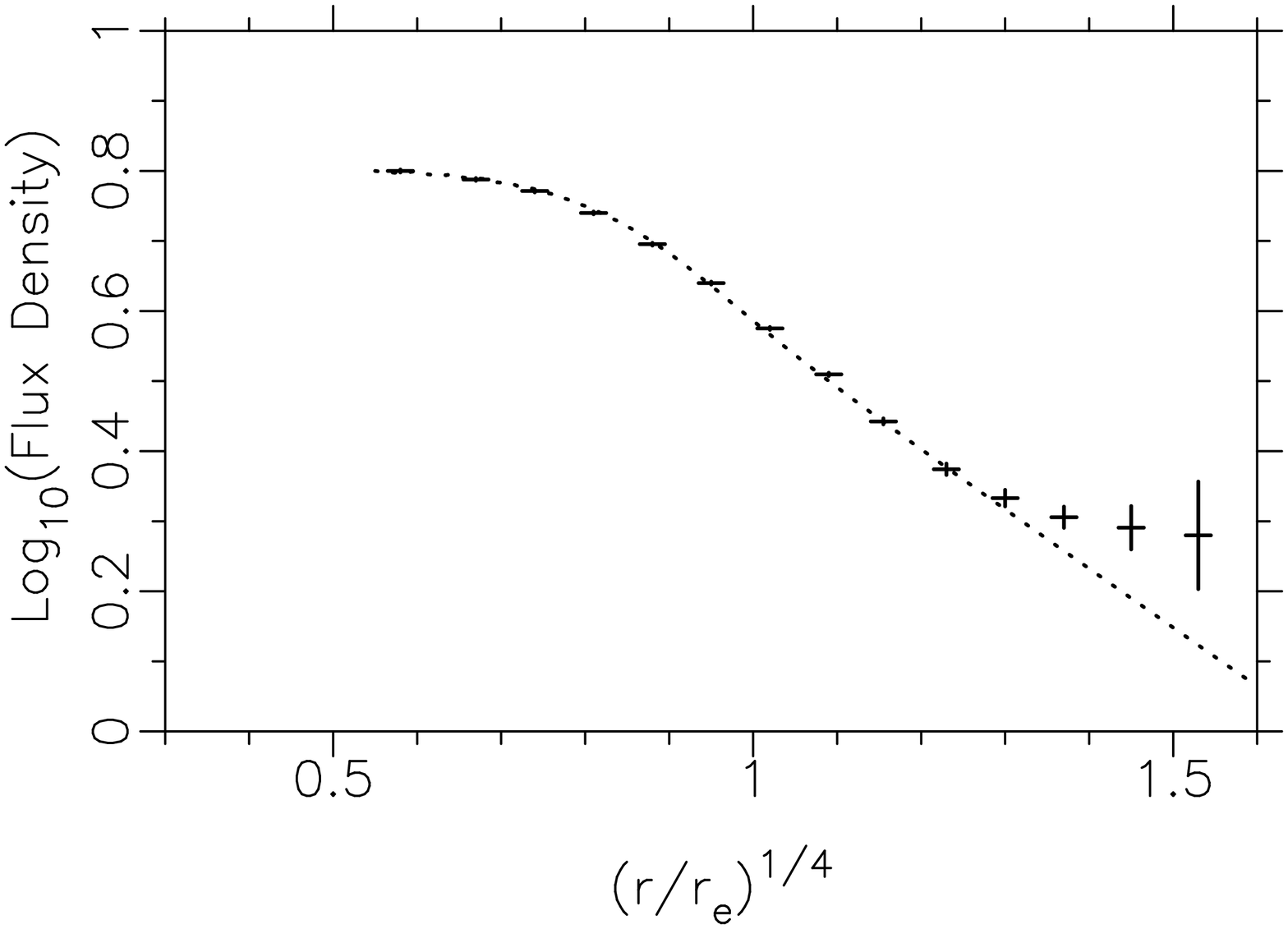,angle=-90,clip=,width=7.8cm}
}
\caption{\label{vaucsum} A combined radial intensity profile in the
K--band for the 12 galaxies with $1.0 \le r_{\rm e} \le 2.0$. Units on the
y--axis are arbitrary. The combined errors for each point are shown. The
dotted line shows a combined de Vaucouleurs profile: halo emission is
clearly visible at $r \gta 2.5r_{\rm e}$ (ie. $(r/r_{\rm e})^{1/4} \gta
1.25$).}
\end{figure}

For a significant fraction of the galaxies, the observed profiles become
brighter than the predicted de Vaucouleurs profile at large radii,
suggesting that at least some of these galaxies possess diffuse extended
envelopes.  The reality of these envelopes for any individual galaxy
cannot be gauged because the signal--to--noise ratio of this extended
emission is low and, more importantly, the errors at large radii are
dominated by uncertainties in the background subtraction and so are not
necessarily independent. We have therefore combined the radial intensity
profiles of the galaxies.  The radial profile of each galaxy was scaled to
the same characteristic radius enabling the galaxy profiles to be summed
directly. Only those galaxies for which $1.0 \le r_{\rm e} \le 2.0$ were
included: for galaxies with a smaller $r_{\rm e}$ there remains a
significant contribution due to the seeing at radii of 2 to 3\,$r_{\rm
e}$, whilst for galaxies with $r_{\rm e}$ greater than 2, there is
insufficient signal at the largest radii to test for the presence of the
halo. The scaled radial profiles of the remaining 12 galaxies were summed,
weighting each galaxy equally, and the results are presented in
Figure~\ref{vaucsum}. 

At radii greater than about 2.4\,$r_{\rm e}$, that is, $\left (r/r_{\rm e}
\right )^{1/4} \sim 1.25$, at which the effects of seeing should be small,
a significant halo component has been detected. This envelope extends from
a physical radius of typically about 35\,kpc out to at least 60\,kpc, and
probably further. This extended envelope may be dominated by the emission
from only a small number of the radio galaxies, but it must be present in
at least some of them.

The halo emission is not an artefact associated with the flat spectrum
contribution to the K--band emission, for two reasons. Firstly, on the HST
images (Paper 1), well over 95\% of the flat spectrum emission originates
at radii less than 35\,kpc, the radius at which evidence for the halo
component first appears; secondly, if the 12 galaxies are split into two
groups according to the percentage of flat spectrum light in the K--band
image, there is no significant difference between the strength of the halo
component in the two samples. We have also studied the profiles of stars
of the images to check that this halo is not an artefact associated with
wings in the UKIRT point spread function.

Such a halo component is typical of cD galaxies (e.g. Oemler
1976)\nocite{oem76}. Thuan and Rominishin \shortcite{thu81} compared the
radial intensity profiles of brightest cluster galaxies in rich and poor
clusters of galaxies, and showed that enhanced halo components were only
seen around the central galaxies of rich clusters, and that they begin at
radii between about 25 and 40\,kpc, comparable to the 35\,kpc derived for
the 3CR galaxies. Figure~\ref{vaucsum} strongly suggests that at least
some of the 3CR galaxies at redshifts $z \gta 1$ are cD galaxies in rich
cluster environments.

\section{Surface photometry and comparison with low redshift elliptical
galaxies} 
\label{remu}

Elliptical galaxies at low redshift possess spectro-photometric properties
which lie on the `fundamental plane', in which their characteristic radii
$r_{\rm e}$, average surface brightnesses within their characteristic
radius $\left<\mu\right>_{\rm e}$, and velocity dispersions, $\sigma_{\rm
v}$, are strongly correlated. The projection of the fundamental plane
using only $r_{\rm e}$ and $\left<\mu\right>_{\rm e}$, known as the
Kormendy relation, also produces a tight correlation and avoids the need
for detailed spectroscopy \cite{kor77,sch87,oeg91}. This provides a
powerful tool for investigating the stellar populations of high redshift
elliptical galaxies.

To compare the 3CR galaxies with existing data for elliptical galaxies at low
redshift, the mean rest--frame V--band surface brightnesses within the de
Vaucouleurs radii are required. Although in many cases an HST image at
rest--frame V--band is available, these are dominated by the flat spectrum
aligned component, and may also be influenced by dust extinction. As shown
above, however, the K--band images are relatively unaffected except in a
few cases; in the analysis that follows we omit 3C22 and 3C41 because of
the presence of nuclear components which contribute to their K--band
emission. 

We adopt the same assumptions as in Section~\ref{iralign}, namely that the
3CR radio galaxies are passively evolving giant elliptical galaxies which
formed at a large redshift, with an additional flat spectrum aligned
component. We use the results of Section~\ref{iralign} to remove the
contributions of any flat spectrum or nuclear component from the K--band
flux densities of the galaxies. The K--band surface brightnesses thus
derived are listed in Table~\ref{irtab2}. The Bruzual and Charlot
\shortcite{bru93} stellar synthesis codes were used to produce the
SED of a passively evolving elliptical galaxy, which formed at $z_{\rm f}
= 10$, at the redshift of each 3CR source. These SEDs were then used to
derive appropriate K--corrections and hence, taking account of the
$(1+z)^4$ cosmological surface brightness dimming, the mean rest--frame
V--band surface brightnesses. The variation of the V--band flux density
with age for each synthesised galaxy was determined, and was used to
convert the derived V--band surface brightness into that which would be
observed when the galaxy had continued to evolve passively through to zero
redshift. This procedure resulted in surface brightnesses of the 3CR
galaxies which can be compared directly with low redshift giant
ellipticals in the fundamental plane.

\begin{figure*}
\centerline{
\psfig{figure=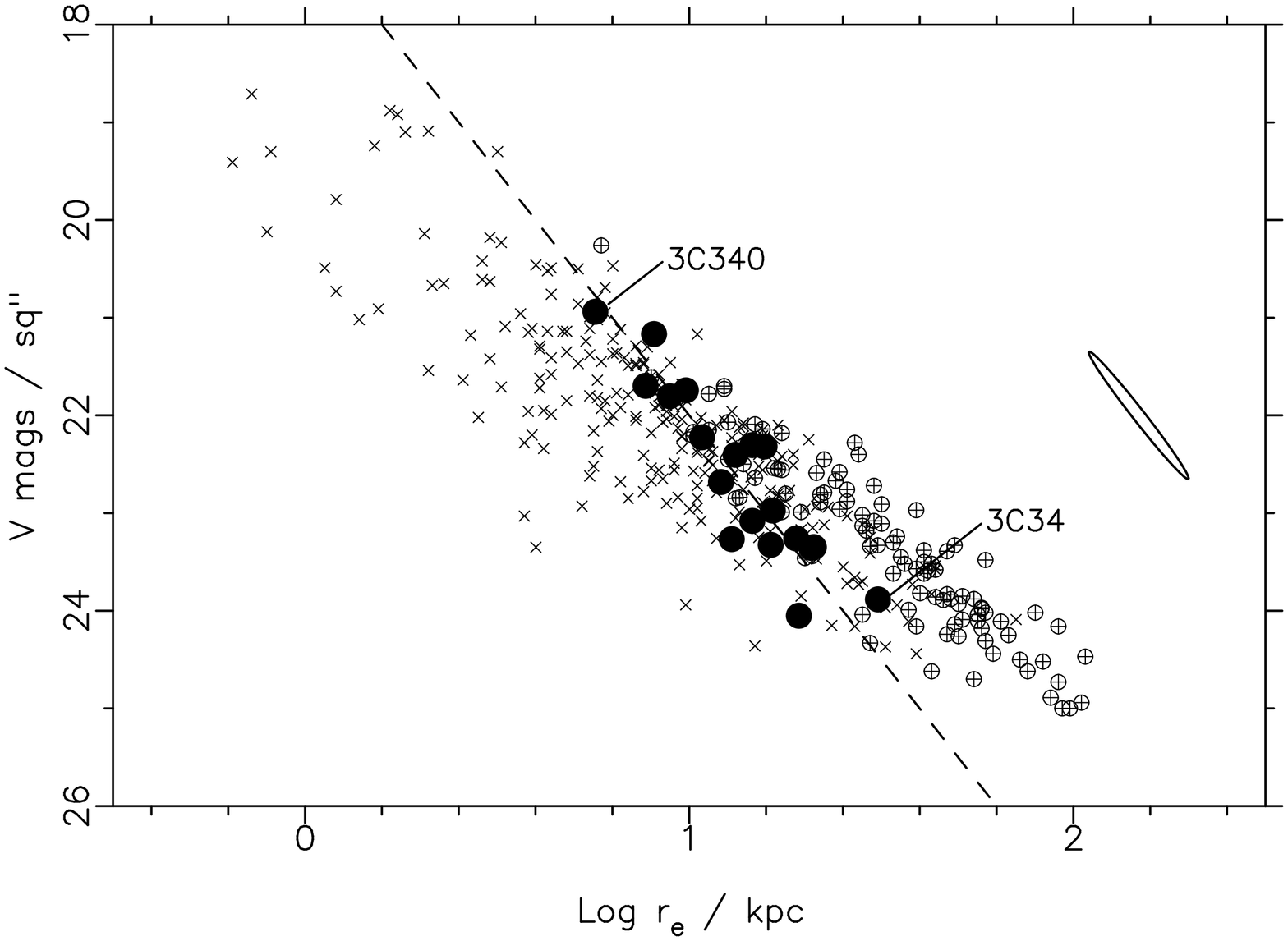,angle=-90,clip=,width=12cm}}
\caption{\label{revsmu} The V--band Kormendy relation for the 3CR galaxies
(solid circles) compared with low redshift ellipticals (crosses) and
brightest cluster galaxies (crossed circles). The low redshift data are
from Schombert (1987). The V-magnitude data of the 3CR galaxies refer only
to the stellar components, and have been derived from their K--magnitudes
assuming that the stellar populations evolve passively.  The ellipse
indicates the error ellipse for the 3CR galaxies for which characteristic
radii were measured from the UKIRT data; those for which the measurements
were obtained from the HST data have error ellipses less than half this
size.  The dashed line shows a line of constant total luminosity.}
\end{figure*}

The derived surface brightnesses are included in Table~\ref{irtab2}, and
are plotted against the de Vaucouleurs radius in Figure~\ref{revsmu}
(solid circles), together with data from low redshift samples of brightest
cluster galaxies in Abell clusters (crossed circles) and other ellipticals
(crosses) taken from Schombert \shortcite{sch87}. The radio galaxies lie
along the fundamental plane defined by the low redshift ellipticals,
providing direct evidence that their stellar populations indeed formed at
large redshifts and are passively evolving. The SEDs, and hence the
K--corrections of either young or non-evolving stellar populations would
be significantly different and the 3CR galaxies would not lie on the
Kormendy relation.

There is some evidence that the slope of the Kormendy relation defined by
the 3CR galaxies is slightly steeper than that of the low redshift
ellipticals.  The dashed line on Figure~\ref{revsmu}b shows a line of
constant total luminosity, that is, a line along which the product of the
surface brightness and the square of the characteristic radius is
constant. The 3CR galaxies lie roughly along such a
line suggesting that, if they all evolved passively until a redshift of
zero, the resultant galaxies would have similar luminosities. This result
is similar to that found by Lilly and Longair (1984) and the inclusion of
the surface brightness information in Figure~\ref{revsmu} confirms that the radio galaxies at $z \sim 1$ are old giant elliptical
galaxies. It seems that the powerful 3CR radio sources are only formed if
their host galaxies attain a specific well--defined stellar mass. We
return to this point in Section~\ref{discuss}.

\begin{figure}
\centerline{
\psfig{figure=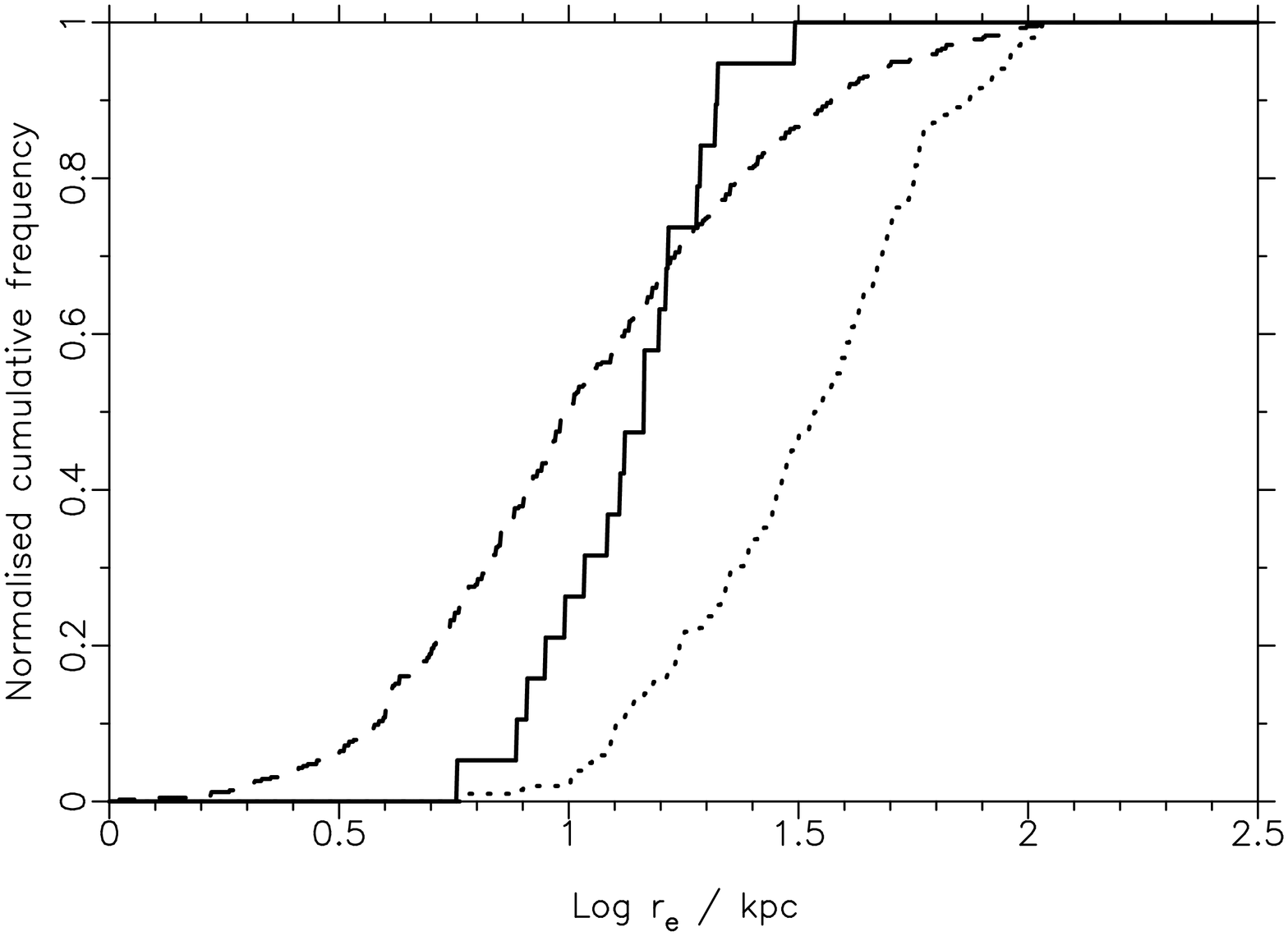,clip=,angle=-90,width=7.8cm}
}
\caption{\label{cumprobre} Normalised cumulative frequency distributions
of $r_{\rm e}$ for the three different populations of galaxies: solid line
-- the high redshift 3CR galaxies; dotted line -- low redshift brightest
cluster galaxies; dashed line -- low redshift elliptical galaxies
(excluding brightest cluster galaxies). The 3CR galaxies have a much
smaller spread in $r_{\rm e}$ than the low redshift elliptical galaxies.}
\end{figure}

The data in Figure~\ref{revsmu} can be used to compare the range of values
of $r_{\rm e}$ of the high redshift 3CR galaxies with those of the
populations of low redshift brightest cluster galaxies and of other
ellipticals. Normalised cumulative frequency distributions of $r_{\rm e}$
for each of the three samples in Figure~\ref{revsmu} are plotted in
Figure~\ref{cumprobre}. Two differences are apparent: firstly, the
dispersion of $r_{\rm e}$ in the 3CR galaxies is significantly smaller
than that of the other two populations; secondly, the mean characteristic
radius of the high redshift 3CR galaxies ($\overline{r_{\rm e}} = 14.6 \pm
1.4$)\,kpc is greater than that of the low redshift ellipticals
($\overline{r_{\rm e}} = 8.2 \pm 1.0$)\,kpc, but about a factor of 2.25
smaller than the low redshift brightest cluster galaxies
($\overline{r_{\rm e}} = 32.7 \pm 1.1$)\,kpc. It should be noted that the
characteristic sizes of the distant 3CR radio galaxies are dependent upon
the adopted value of $q_0$ and, for $q_0 = 0$, would be approximately 25\%
smaller.  On the other hand, the small degeneracy between the detection of a
point source component and the value of the characteristic radius,
discussed in Section~\ref{vaucpro}, means that if there were to be a small
nuclear contribution to these galaxies then the characteristic radii would
be up to 20\% larger than those derived. Neither of these effects would
change the significance of the results.

According to cannibalism models (e.g. Hausman and Ostriker
1978)\nocite{hau78} the position of a galaxy along the fundamental plane
is interpreted as being related to its merger history. Assuming that the
mergers occur homologously (e.g. Schombert 1987\nocite{sch87} and
references therein), when a small galaxy is cannibalised by a much larger
galaxy to produce a bound remnant, a large fraction of the kinetic energy
of the merger is converted into binding energy of the remnant. The result
of such a homologous merger is that the remnant has a larger radius and
more diffuse morphology as compared with the original galaxy, thus moving
the galaxy to the right along the projected fundamental plane in
Figure~\ref{revsmu}\ \cite{hau78}. In this respect, the larger size of the
high redshift 3CR galaxies as compared to low redshift elliptical galaxies
is important: despite being observed at a much earlier cosmic epoch, the
3CR galaxies at $z \sim 1$ appear to be highly evolved dynamically. We
discuss the relative sizes of the distant 3CR galaxies and the brightest
cluster galaxies in Section~\ref{discuss}.

Within the 3CR sample, it is interesting that 3C34 has the largest
characteristic radius, suggesting that it has undergone most mergers,
whilst 3C340 is the most compact object. The former of these is known to
lie towards the centre of a moderately rich cluster of galaxies
\cite{mcc88,bes97b}, whilst there is practically no depolarisation of the
radio lobes of the latter \cite{joh95} suggesting that there is relatively
little cluster gas surrounding the galaxy.

\section{The revised K$-z$ diagram}
\label{seckzdiag}

Much evidence has been presented in the previous sections confirming the
stellar nature of the infrared emission of the galaxies that host the 3CR
radio sources. A revised K$-z$ relation for the 3CR galaxies has been
derived, including a number of improvements over the original version
presented by Lilly and Longair \shortcite{lil84a}:

\begin{figure}
\centerline{
\psfig{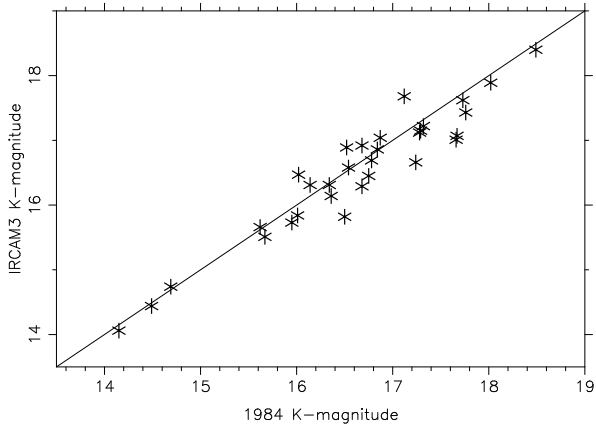}
}
\caption{\label{oldvsnew} A plot of the K--magnitudes measured for the 28
3CR sources in the recent IRCAM3 observations against those of Lilly and
Longair (1984). In the diagram, the new K magnitudes were reduced to the
same apertures used by Lilly and Longair in their single element
photometry. Most of the discrepancies are within the quoted errors on the
original observations, but there are a few significant changes.}
\end{figure}

\begin{enumerate}
\item The K--magnitudes measured by Lilly and Longair were obtained by
single element aperture photometry using blind offsets and chopping. The
new magnitudes show that, although in the majority of cases the 1984
infrared magnitudes were remarkably accurate, in a small number of cases
the on--source apertures were sufficiently large to contain emission from
companions, or the off--source reference beam included a faint companion
(Figure~\ref{oldvsnew}).  The new imaging infrared observations allow
these problems to be eliminated.

\item Lilly and Longair had to make significant aperture corrections for
many of the high redshift radio galaxies; in most cases, our magnitudes
were measured through the adopted 9 arcsec diameter aperture. Any extended
halos are generally only present at a radius $r \gta 35$\,kpc (4 arcsec at
$z \sim 1$), and so the contribution of any halo to the total K--band flux
density within the 9 arcsec diameter aperture adopted for our photometry
is unimportant.

\item The sources 3C22 and 3C41 are not included with the narrow--line
galaxies, but are classified with the broad line galaxies in which a
nuclear component makes a significant contribution to the K--band flux
density. 

\item The contribution of the aligned component to the K--band magnitude
has been removed by subtracting the appropriate fraction of flux density
as estimated by our SED fitting in Section~\ref{fitspecs}. The fraction of
the total flux density that might be associated with any point source
component has also been removed. The combination of these two corrections
generally amounts to less than 20\%, corresponding to about 0.2 magnitudes.
The corrected K$_{\rm corr}$ magnitudes are listed in Table~\ref{irtab1}.
\end{enumerate}

\begin{figure*}
\centerline{
\psfig{figure=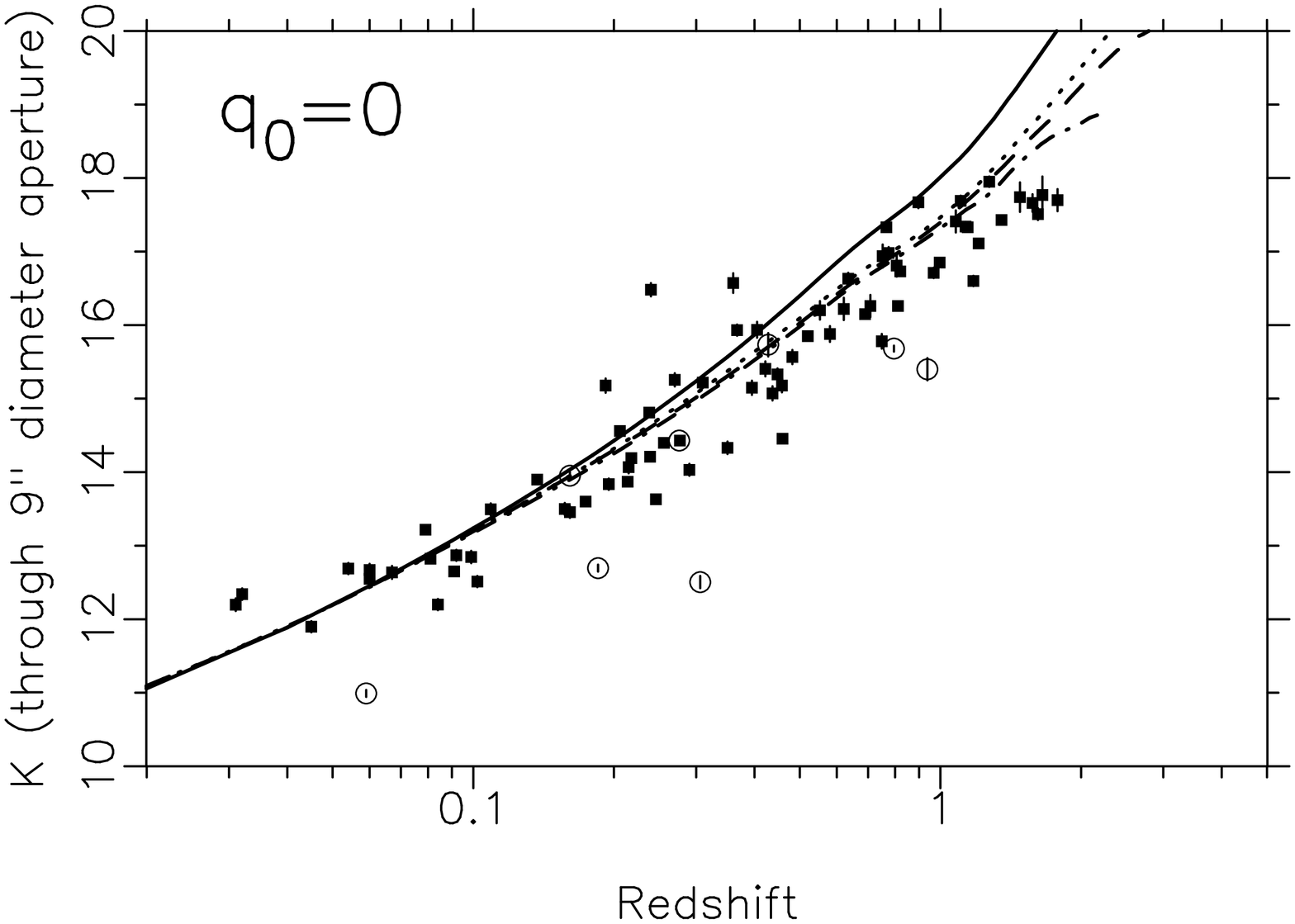,clip=,width=12cm,angle=-90}
}
\medskip
\centerline{
\psfig{figure=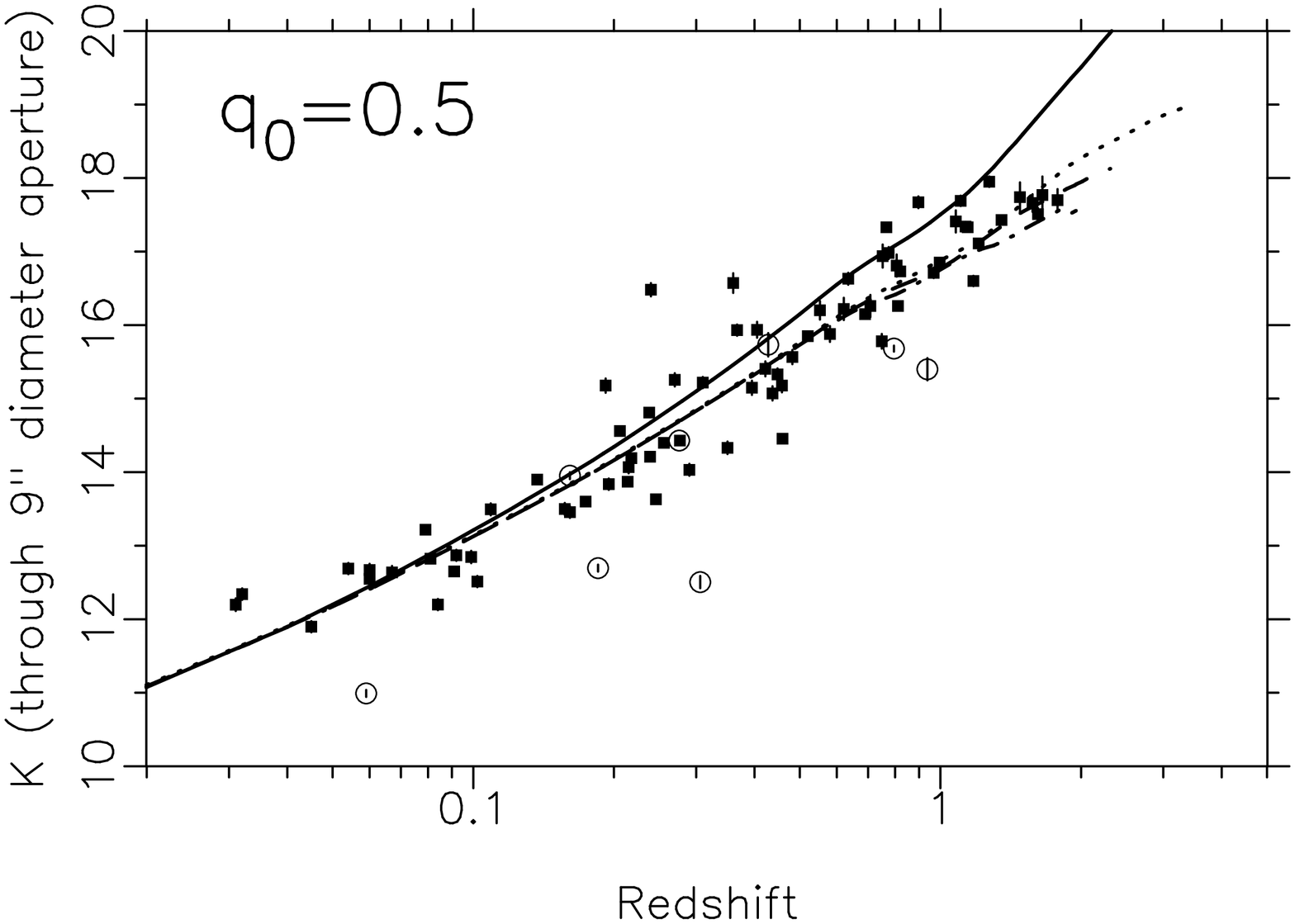,clip=,width=12cm,angle=-90}
}
\caption{\label{kzdiag} Plots of K--magnitude {\it vs} redshift for the
3CR radio galaxies. Broad line radio galaxies are marked by open circles
as are 3C22 and 3C41, and narrow line radio galaxies by filled
squares. The error in the K--magnitude measurements is indicated by the
vertical error bar associated with each point; in many cases, this is
smaller than the plotted symbol. The upper plot corresponds to $q_0 = 0$
and the lower plot to $q_0 = 0.5$. The lines on each plot show the
predicted K$-z$ tracks of various evolutionary models and have been
reduced to the magnitudes observed within a 9 arcsec diameter aperture at
each redshift.  The solid lines correspond to a non--evolving stellar
population, whilst the dotted, dashed, and dash--dot lines are for
passively evolving galaxies formed at $z_{\rm f} = $20, 5 and 3
respectively. The theoretical curves are normalised to match the low
redshift ($z \lta 0.1$) narrow--line radio galaxies.}
\end{figure*}

For the radio galaxies with redshifts $z < 0.6$, the photometry of Lilly and
Longair \shortcite{lil84a} was adopted.  These galaxies are sufficiently
bright that offsetting onto nearby faint companions would have had only a
small effect upon the observed magnitudes, and there is no evidence for a
strong alignment effect at low redshifts. The magnitudes of these galaxies
have all been scaled to a 9 arcsec diameter aperture, using a radial light
profile appropriate for low redshift radio galaxies (e.g. Lilly \etal\
1984b):\nocite{lil84c} the corrections required are small, being less than
$0.3$ magnitudes in all cases, and generally less than $0.1$ magnitudes.

The revised K$-z$ relation is shown in Figure~\ref{kzdiag}. The
broad--line radio galaxies, together with 3C22 and 3C41, are plotted as
open circles. The predicted K$-z$ tracks for various evolution models and
cosmological parameters are also shown, normalised to match the low
redshift ($z \lta 0.1$) narrow--line radio galaxies. These tracks include
a correction based upon a standard radial light curve for low redshift
giant elliptical galaxies, to account for the fact that the galaxies are
observed through an aperture of fixed angular size of 9 arcsec, rather
than fixed physical size. At redshifts $z \sim 1$, generally over 90\% of
the flux density is contained within this aperture.

Figure~\ref{kzdiag}a shows a set of tracks for a $q_0 = 0$ universe,
whilst Figure~\ref{kzdiag}b shows the corresponding tracks in a $q_0 =
0.5$ universe, in both cases the cosmological constant $\Lambda$ being
taken to be zero. In each case, the solid line represents the track of a
non--evolving giant elliptical galaxy, obtained by redshifting the spectrum
of a giant elliptical galaxy at the present epoch. The dashed, dotted,
dash--dot lines show the predicted tracks of passively evolving galaxies
formed in a 1~Gyr burst at three different redshifts. A good fit to the
data is obtained by the passively evolving models in the $q_0 = 0.5$
Universe.  Little difference is found if the galaxies formed at redshift
$z = 5$ or $z = 20$, but at redshifts $z \gta 1.2$ the predicted track for
galaxy formation at $z = 3$ is significantly brighter than the observed
data, indicating that the stellar populations formed at an earlier cosmic
epoch. None of the tracks in the $q_0 = 0$ plot provide a good fit to the
observations.  

\section{Discussion}
\label{discuss}

\subsection{The cluster environments of the distant 3CR radio galaxies}
\label{clusters}

The amplitudes of the spatial cross--correlation functions for galaxies in
the vicinity of nearby 3CR FR\,II class radio sources \cite{fan74} are
similar to those for normal elliptical galaxies \cite{pre88}, whilst the
host radio galaxies themselves have optical luminosities and
characteristic sizes significantly less than those of first ranked Abell
cluster galaxies \cite{lil87}. This indicates that low redshift FR\,II
sources typically lie in isolated environments or small groups. In
contrast, measures of the galaxy cross--correlation function \cite{yat89}
and an Abell clustering classification \cite{hil91} both indicate 3CR
galaxies with intermediate redshifts, $0.3 \lta z \lta 0.5$, belong to
environments which are about three times richer than those nearby.

There is evidence that the high redshift 3CR sources belong to moderate to
rich cluster environments, in particular: the detection of X--ray
emission, in some cases resolved, from a number of these sources has been
associated with a cooling flow in relatively dense intracluster medium
\cite{cra93,wor94,cra95,cra96b,dic97a}; companion galaxies are seen around
a number of the radio galaxies in the narrow--line [OII]~3727 imaging by
McCarthy \shortcite{mcc88}; infrared imaging of distant radio galaxies
provides large numbers of cluster candidates, based upon selection by
colour, and spectroscopic follow--up of individual 3CR radio galaxies
indeed shows them to lie in at least moderately rich clusters
(e.g. Dickinson 1997);\nocite{dic97a} a surrounding medium of fairly high
density is required to account for the large rotation measures of distant
radio sources, and to provide a working surface for their high radio
luminosities \cite{car97,bes97i}.

In this paper we have provided further evidence in support of this
hypothesis. In Section~\ref{vaucpro} we showed that at least some of these
galaxies possess halos characteristic of those seen around cD galaxies at
low redshifts. Furthermore, the stellar masses of the 3CR galaxies derived
in Section~\ref{fitspecs} indicate that they are amongst the most massive
galaxies at cosmic epochs at which they are observed.  A comparison of the
3CR galaxies with the brightest cluster galaxy (BCG) sample of
Arag{\'o}n--Salamanca \etal\ \shortcite{ara93} shows that the absolute
K--magnitudes, and hence stellar masses, of the two samples at redshift $z
\sim 1$ are essentially the same. In contrast, at low redshifts, BCGs are
about a magnitude brighter than the 3CR radio galaxies in absolute
K--magnitude, consistent with the result that low redshift 3CR galaxies
lie in isolated environments of small groups. This difference is best
illustrated by the fact that the K$-z$ relation of the BCGs is consistent
with non-evolving stellar populations \cite{ara93}.

To understand the K$-z$ relations, we must therefore unravel
two `cosmic conspiracies':

\begin{enumerate} 
\item On the one hand, the tightness and slope of the 3CR K$-z$ relation
suggests that the radio galaxies belong to a single population of objects
which formed at large redshift and evolved passively over the redshift
interval $2 > z > 0$, such that a high redshift 3CR galaxy would evolve
into a low redshift 3CR galaxy. This simple picture cannot be correct
because we have argued that the galactic environments of FR\,II 3CR radio
galaxies change with redshift. High redshift 3CR galaxies appear to lie in
(proto--)cluster environments, whilst the nearby sources in the sample are
found in isolated environments or small groups.

\item On the other hand, the K$-z$ relation of the BCGs suggests that
their stellar populations are non--evolving, but this is clearly an
unphysical picture. All passively evolving models for stellar
populations suggest that they should be about a magnitude brighter in
absolute K--magnitude at redshift $z = 1$ as compared with $z = 0$.
\end{enumerate}

\begin{figure}
\centerline{
\psfig{figure=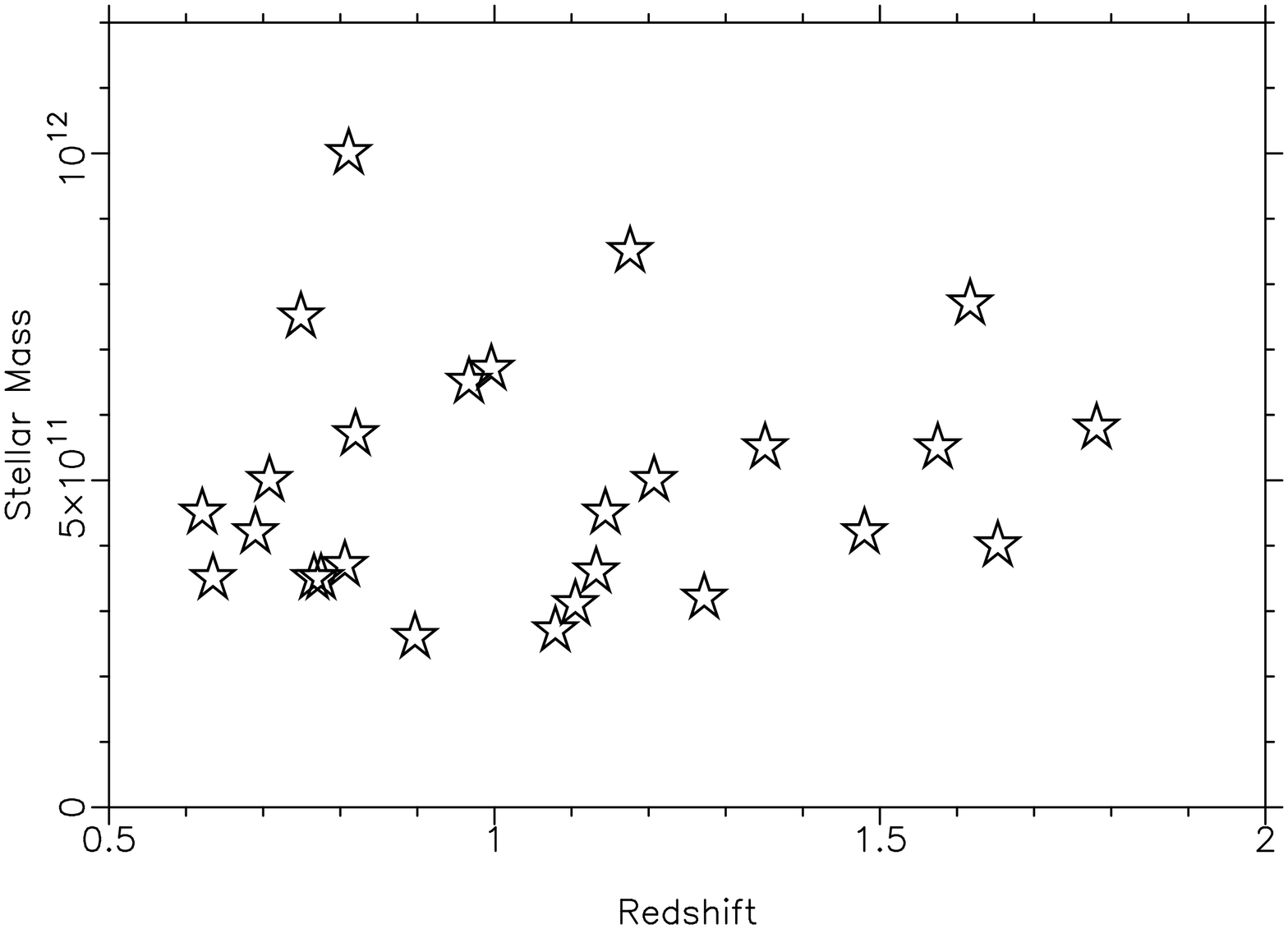,angle=-90,clip=,width=7.8cm}}
\caption{\label{massz} A plot of the stellar mass derived by a fit to the
broad band photometry of Section~\ref{iralign} against redshift for 26 3CR
galaxies.}
\end{figure}

Since the stellar populations of the BCGs must have evolved in the same
way as the 3CR radio galaxies, the shape of their K$-z$ relation reflects
the fact that central cluster galaxies continue to accumulate matter
through mergers and gas infall.  Hierarchical clustering models for
structure formation have suggested that the mass of BCGs can increase by a
factor of up to 5 between a redshift of one and the present epoch
\cite{ara97}. In contrast, the variation of the stellar masses of the 3CR
galaxies with redshift, as derived from the SED fits of Section 3.1, shows
no correlation (see Figure~\ref{massz}).

In Section~\ref{remu} we showed that the mean characteristic radius of the
distant 3CR radio galaxies was approximately a factor of 2.25 smaller than
that of the nearby BCGs. Kormendy \shortcite{kor77} derived a relation
between the integrated luminosity of a bright elliptical galaxy and its
characteristic radius: $L_{\rm int} \propto r_{\rm e}^{0.7}$, or
equivalently, $M_{\rm stel} \propto (M/L)_0 r_{\rm e}^{0.7}$, where
$M_{\rm stel}$ is the stellar mass of the galaxies and $(M/L)_0$ is the
mass--to--light ratio of the stellar populations at redshift zero. This
result was confirmed for elliptical galaxies with $r_{\rm e} \gta 6$\,kpc
by Schombert \shortcite{sch87}. This implies that the stellar masses of
the 3CR radio galaxies at redshift $z \sim 1$ are only a factor of about
1.75 below those of nearby central galaxies in Abell clusters, assuming
that their stellar populations will have a similar mass--to--light ratio
when they have evolved to redshift zero. When compared to the growth
factor of between 3 and 5 expected for BCGs over this redshift interval, it
is clear that the distant 3CR galaxies are massive galaxies, even relative
to the progenitors of the central galaxies of rich cluster.

Intriguingly, West \shortcite{wes94} has shown that the comoving
number density of powerful high redshift radio galaxies is of the same
order of magnitude as that of rich Abell clusters, if account is taken of
the fact that radio source lifetimes, $\sim 10^8$ years, are much shorter
than the age of the Universe. It is quite conceivable that all of the
brightest galaxies in clusters today went through a phase of during which
they were intense radio sources, and so must possess supermassive black
holes in their nuclei (e.g. van der Marel \etal\ 1997 and references
therein).\nocite{mar97}

\subsection{Understanding the 3CR radio galaxies}
\label{3crgals}

Since 3CR radio galaxies at large redshifts lie at the centre of galaxy
clusters, their masses would be expected to increase with cosmic epoch
just like the BCGs; the `passively evolving' K$-z$ relation of these
galaxies suggests, however, that this does not occur. This is a
conspiracy: the galaxies sampled at high and low redshifts do not form a
uniform population, as is indicated by the dramatic change in their
galactic environments with redshift. To understand this effect, we compare
the K$-z$ relation of the 3CR galaxies with that of the only other
complete sample of low--frequency selected radio sources, the 6C radio
galaxies \cite{eal96,eal97}. There are important differences between the
two samples, specifically:

\begin{enumerate}

\item At a given redshift, the 3CR radio sources are about a factor of 5
more powerful than the 6C radio sources, since the latter are selected
from a fainter radio survey.

\item At low redshifts, $z \lta 0.6$, the 3CR and 6C radio galaxies have
similar infrared luminosities.

\item At redshifts $z \sim 1$, the (un-corrected) K--band luminosities of
the 3CR galaxies are greater than those of the 6C galaxies by, on average,
0.6 magnitudes \cite{eal97} although there is significant overlap between
the two populations.

\end{enumerate}

\noindent Points (ii) and (iii) reflect the fact that the
K$-z$ relation for 6C radio galaxies is consistent with a non--evolving
population of radio galaxies \cite{eal96}.

Three possible reasons why the K--magnitudes of the 3CR galaxies might be
brighter than those of the 6C galaxies at $z \gta 1$ are: (i) there may be
a strong direct or indirect AGN contribution to the K--band emission of
the high redshift 3CR galaxies (e.g. Eales \etal\ 1997);\nocite{eal97}
(ii) the 3CR galaxies may be younger than the 6C galaxies, and hence their
stellar populations would be more luminous; (iii) the 3CR galaxies may be more
massive than the 6C galaxies and thus contain a greater number of stars.

Concerning the first possibility, nearly 50\% of the K--band emission of
the 3CR galaxies would have to be associated with AGN activity. The
results presented in this paper show that, with the exceptions of the two
sources 3C22 and 3C41, the mean point source contribution to the total
K--band flux density is only 4\%, with secure upper limits of $\lta
15\%$. We have discussed why line emission from gas excited by the AGN
light is unlikely to be important, and have estimated the mean
contribution of the aligned emission to the K--band flux density to be
about 10\%. We conclude, therefore, that the sum of these effects will
contribute at most about 0.3 of the 0.6 magnitudes difference
between the two samples.

In the second case, if the 3CR galaxies were formed at a smaller redshift
than the 6C galaxies, say $z_{\rm f} \sim 2.5$, their younger stellar
populations would be more luminous than the corresponding 6C galaxies at
redshifts $z \gta 1$, but by $z \sim 0.6$ they would have aged and the
galaxies would no longer be significantly more luminous. The main
objections to this picture are that features detected in the spectra of
some of the galaxies, coupled with their very red infrared colours, imply
much older ages for the stellar populations of the 3CR radio galaxies than
would be expected in this picture (eg. Stockton \etal\
1995)\nocite{sto95}, and that it is difficult to assemble such massive,
dynamically evolved systems as the 3CR galaxies in a short cosmic time.

It is likely, therefore, that both the 6C and 3CR populations formed at
large redshifts, and that the 6C galaxies are fainter than the 3CR galaxies
in the K--band at high redshift because they contain, on average, a lower
mass of stars. The primary selection criteria for these samples of radio
galaxies are the limiting radio flux densities of the 3CR and 6C
catalogues, and so the different K$-z$ relations of these two samples
imply that the radio luminosities of the sources must depend upon the
masses and environments of the host galaxies.

According to conventional ideas about the origin of FR\,II radio sources,
their luminosities are ultimately determined by the powers of the beams of
particles ejected by the AGN, and by the density of the surrounding
interstellar and intergalactic gas which controls adiabatic radiation
losses of the radio lobes. In turn, the intrinsic beam power of the source
is determined by the mass of the central black hole, the mass of material
accreting onto it, and the efficiency with which this is converted into
beam energy. Given an abundant supply of fuel, it is a reasonable
assumption that the total energy output of the central engine, which is
dominated by the bulk kinetic power of the radio jets \cite{raw91b}, will
be close to the Eddington limiting luminosity.  Rawlings and Saunders
derived the total kinetic power of the jets in radio galaxies with
redshifts $z \sim 1$ and found values corresponding to the Eddington
limiting luminosity of a black hole with a mass of a few times $10^8
M_{\odot}$; Kaiser \etal\ \shortcite{kai97b} suggest that the jet powers
may be even higher. Theorists have long argued that the masses of the
central black holes should be roughly proportional to the masses of the
host galaxies \cite{sol82,efs88,sma92,hae93}, and some evidence for such a
correlation has recently been found for massive black holes in the nuclei
of nearby galaxies \cite{kor95}. Thus, the luminosity of a radio source is
determined by three principal factors: the mass of the host galaxy, the
availability of fuel, and the density of the environment.

The gas which forms the fuel for radio sources may arise either through
the infall of interstellar and intergalactic material towards the central
regions of the galaxy, or as a result of an interaction or merger with a
gaseous dwarf galaxy \cite{hec86,smi89,smi90,men92}. In high redshift
proto--cluster or young cluster environments, the central galaxies will be
massive, and so contain amongst the most powerful central engines at that
epoch.  There will also be high merger rates and a plentiful supply of
disturbed intracluster gas to fuel the central engine and confine the
radio lobes. Furthermore, Ellingson \etal\ \shortcite{ell91} have found
that the velocity dispersions of galaxies around distant powerful radio
sources are significantly lower (400 to 500\,km\,s$^{-1}$) than those in
comparably rich clusters at low redshifts (500 to 1000\,km\,s$^{-1}$).
These smaller relative velocities increase the efficiency of galaxy
mergers. Thus, massive galaxies in young cluster environments at high
redshifts possess all the necessary ingredients for producing the most
powerful radio sources.

Radio sources in galaxies of lower mass than the 3CR galaxies are likely
to be powered by less massive black holes and therefore to have lower
intrinsic beam powers. It is reasonable to attribute the lower radio
luminosity of the high redshift 6C radio sources as compared with the 3CR
sources to the fact that the 6C galaxies are less massive and possess less
massive black holes. The overlap between the two samples on the K$-z$
relation would then arise from the scatter in the correlation between the
stellar mass of the galaxy and the mass of the black hole (e.g. Kormendy
and Richstone 1995)\nocite{kor95}; for example, a galaxy with a
K--magnitude comparable to that of the 3CR galaxies, but whose central
black hole is less massive than predicted by the correlation, would host a
6C radio source. 

At redshifts $z < 0.6$, the lower radio luminosity of all radio sources
means that the galaxies are no longer producing radio beams with kinetic
powers close to the Eddington limit. The analysis of Rawlings and Saunders
\shortcite{raw91b} confirms that the beam powers of these sources are
significantly lower than the values found in the radio galaxies at $z \sim
1$. Neither the masses of individual black holes, nor the galaxy
clustering can decrease with cosmic epoch, and so the radio luminosities
of low redshift sources must be limited not by the mass of the host
galaxy, as are high redshift sources, but rather by the availability of
gas in the host galaxy and its environment to fuel the central black
hole. As a result, at low redshift a weaker correlation would be expected
between galaxy mass and radio luminosity.  This may account for the
similarity of the K--band magnitudes of the 3CR and 6C galaxies at these
redshifts. 

This decrease in the fuelling gas density with cosmic epoch has been
argued on different grounds by other authors (e.g. Rees 1990
\nocite{ree90} and references therein). It is also consistent with recent
results on the evolution of the fraction of baryons in the form of neutral
hydrogen, and the enrichment of the heavy elements, which indicate that
the quantity of interstellar and intracluster material has decreased
significantly from redshifts $z \sim 2$ to $z = 0$ \cite{mad96}.

\subsection{Cosmic evolution of the radio source population} 
\label{radevol}

We have argued that radio galaxies at high redshifts lie in young cluster
environments, and that such environments possess the ideal ingredients for
producing powerful radio sources. Why then are the most powerful FR\,II
radio sources at the present epoch not also associated with central
cluster galaxies?  The radio sources hosted by nearby BCGs are almost
invariably `edge--darkened' FR\,I type radio sources. FR\,Is do not
possess large luminous radio lobes; instead, the jets are decelerated
significantly in the inner kiloparsec, probably due to entrainment of
external material (Laing 1993 and references therein)\nocite{lai93},
implying a relatively low kinetic power for FR\,I jets. How do the
environments of low redshift central cluster galaxies restrict, in some
way, the central engines of these galaxies to producing only the lower
power FR\,I jets?

Narayan and Yi \shortcite{nar95} have shown that if the accretion rate on
to a black hole falls below a critical value given by $\dot{M}_{\rm crit}
\approx \alpha^2 \dot{M}_{\rm Edd}$ (where $\alpha$ is the viscosity
parameter which has a value of order 0.1) then accretion on to the black
hole occurs in an advection dominated mode. At higher accretion rates a
radiatively efficient thin accretion disk forms. Studying the core spectra
of the radio loud quasar 3C273 and the FR\,I source M87, Reynolds \etal\
\shortcite{rey96} argue that the difference between FR\,I and FR\,II
sources may be that the central engines of FR\,I sources accrete matter at a
rate below the critical value.  The large difference in jet luminosity
between FR\,I and FR\,II type sources would then arise from a much smaller
difference in the mass accretion rate, coupled with a change in the
accretion mechanism.

Baum \etal\ \shortcite{bau92} showed that the gas surrounding FR\,I radio
galaxies is of low angular momentum, with no evidence for recent mergers,
whilst that surrounding FR\,II sources is generally of higher angular
momentum and frequently exhibits tidal tails or other evidence for
mergers. Rich regular clusters at low redshifts have had time to virialise
the spatial distribution of the galaxies, and for the intra\-cluster gas
to take up an equilibrium configuration within the cluster gravitational
potential. Studies of the Butcher--Oemler effect \cite{but78} indicate
that, compared to high redshift clusters, there is a dearth of gas--rich
galaxies close to the centre of low redshift clusters which might merge
with, and fuel, the central galaxy. Furthermore, the high velocity
dispersion of the galaxies greatly reduces the merger efficiency.  The
evolution and virialisation of low redshift clusters is therefore likely
to result in a more restricted fuelling supply for the central engines of
BCGs, and the fuelling gas may arise solely through steady accretion from
the interstellar or intracluster media (see also Baum \etal\
1995\nocite{bau95}). The most powerful FR\,II radio sources at small
redshifts are generally found instead in small groups of galaxies: in
these groups, velocity dispersions are small, mergers relatively frequent,
and gas--rich galaxies plentiful. Occasionally, dramatic mergers will
occur in low redshift BCGs, producing radio sources of comparable radio
power to the $z \sim 1$ 3CR sources, and we ascribe exceptional sources
such as Cygnus A to this process (see also Barthel and Arnaud,
1996)\nocite{bar96}.

These considerations suggest an explanation for the constancy of the
stellar masses of the 3CR galaxies throughout the redshift range $0.03 \le
z \le 1.8$.  This mass may be interpreted as the typical mass which a
giant elliptical galaxy attains before its galactic and gaseous environment
is virialised.  Galaxies of mass greater than a few times $10^{11}
M_{\odot}$ will have undergone greater dynamical evolution, and generally
live in virialised environments where the reduced supply of fuelling gas
to the central regions results in FR\,I sources being formed.  This leads
naturally to a correlation between the redshift of a 3CR FR\,II galaxy and
the richness of its environment: galaxies in highly overdense environments
accumulate matter the fastest, and therefore reach this `FR\,II upper mass
limit' at early cosmic epochs, whilst those which lie in less dense
environments evolve more slowly and can form powerful FR\,II radio
galaxies at the current epoch.

In the model presented here, the location of the distant 3CR galaxies in
the low redshift fundamental plane can be understood. All galaxies evolve
through accumulation of matter and move to the right along the fundamental
plane. For a galaxy to form a radio source at $z \sim 1$ of sufficient
radio power to be in the 3CR sample, the galaxy (and hence its central
black hole) must have attained a certain mass, and therefore a certain
luminosity, which is close to the dashed line on Figure~\ref{revsmu}b. The
small scatter in their characteristic radii follows naturally from the
galaxies being seen at a similar point in their evolutionary history.

This model is also fully consistent with the redshift evolution of the
alignment effect seen in powerful radio galaxies (e.g. McCarthy 1993,
\nocite{mcc93} Paper III). At high redshifts, the sources lie in
environments in which the surrounding gas is unsettled, distributed
throughout the forming cluster, and of relatively high density. Most
models of the alignment effect depend critically upon the availability of
cool gas and so, at intermediate and low redshifts where the environments
of the sources have much lower gas densities and the gas is more settled,
there is a decrease in the luminosity of aligned emission.

These results suggest that the highest redshift ($z \gta 2$) radio
galaxies, like the 3CR galaxies, are amongst the most massive systems at
their observed epoch (see also Pentericci \etal\
1997)\nocite{pen97}. These galaxies must have accumulated their mass at
early cosmic epoch, and so must belong to environments in which the most
massive systems are formed by a redshift of zero. The dense surrounding
environment and the abundant supply of fuel required are consistent with
the detection of large \la\ halos around these objects (e.g. R\"ottgering
and Miley 1996 and references therein)\nocite{rot96d}, and the very large
depolarisation and rotation measures of their radio emission \cite{car97}.

A combination of luminosity evolution of the radio source population, and
the greater availability of gas at high redshifts increasing the
probability of a powerful radio source being formed, provides a natural
explanation for the increase in the comoving number density of radio
galaxies and quasars out to redshifts $z \sim 2$. The levelling off and
decline in the comoving number density of powerful radio sources beyond
this redshifts $z > 2$ can plausibly be explained in terms of the fact that
systems which are rich enough to produce sufficiently massive host
galaxies, and consequently massive enough black holes in their nuclei, had
not had time to form by that epoch. 

At the same time, these results also provide an explanation of why other
populations of active galaxies, such as the optically selected quasars and
the bright X--ray galaxies, show the same evolutionary behaviour as the
powerful radio galaxies \cite{dun94}: since the maximum radio luminosity
is governed by Eddington limited accretion, this same process will
define the limiting optical and X--ray luminosities. Therefore, the cosmic
evolution of all classes of active galaxy are governed by the same two
parameters: the evolution of black holes masses, and the decrease in the
availability of fuelling gas with cosmic epoch.

\section{Conclusions}
\label{concs}

The main conclusions of the study of the old stellar populations of these
3CR galaxies are as follows:

\begin{itemize}

\item The magnitudes, colours and location of the distant 3CR galaxies on
the projected fundamental plane for elliptical galaxies indicate that
their stellar populations formed at large redshift and are passively
evolving. Their spectral energy distributions can be well matched using
such an old stellar population together with an aligned component of
relatively flat spectrum.

\item Except in two cases (3C22 and 3C41), there is no evidence for a
nuclear point source contribution to the K--band emission at $\gta 15\%$
of the total flux density at that wavelength.

\item At least some of the high redshift 3CR galaxies possess extended
envelopes, similar to the halos seen around nearby cD galaxies. Such
envelopes are not seen around low redshift 3CR galaxies, indicating a
difference between the high and low redshift populations. The high
redshift 3CR galaxies appear to live in environments which will form rich
clusters by a redshift of zero.

\item The 3CR radio galaxies have a similar stellar mass over the complete
redshift range of the sample.  The less powerful 6C radio sources have
lower stellar masses at high redshifts; the black holes associated with
these galaxies are likely to be correspondingly less massive, producing
proportionally weaker radio sources.

\item Powerful radio sources are not observed in the most massive galaxies
at low redshift. A reduced supply of fuelling gas for the central engines,
possibly associated with the lower merger rates and lack of gas--rich
galaxies towards the centre of virialised clusters, results in lower
luminosity FR\,I type radio sources being preferentially formed.

\item If the 3CR galaxies evolve in the same way as brightest cluster
galaxies, then the evolution in the comoving number density of powerful
radio sources can be naturally understood: out to redshifts $z \sim 2$,
the comoving number density of powerful radio sources increases through a
combination of luminosity and density evolution associated with the
greater mass accretion rates and the greater probability of a strong radio
source being formed; at redshifts $z \gta 2$, the levelling off and decline
of the comoving number density would be due to negative density evolution,
associated with the decreasing number of galaxies which lie in
environments sufficiently rich enough for them to have accumulated enough
matter to produce a powerful radio source by that epoch.

\item The commonly--used `uniform population, closed box' galaxy evolution
models are not appropriate for interpretations of the K$-z$ relations.
Powerful radio galaxies selected at high and low redshifts have different
evolutionary histories, but must contain a similar mass of stars, a few
times $10^{11} M_{\odot}$, and so conspire to produce the `passively
evolving' K$-z$ relation observed. The apparently simple `no--evolution'
shape of the K$-z$ relation for brightest cluster galaxies and lower power
radio galaxies is the result of a further conspiracy in which these
galaxies become intrinsically fainter due to stellar evolution but also
brighten because of mass increase through mergers. The shapes of these
relations can be used to provide information about the merger histories of
massive galaxies in clusters.
\end{itemize}

\section*{Acknowledgements} 
\label{acknowl}
 
This work is based on observations with the NASA/ESA Hubble Space
Telescope, obtained at the Space Telescope Science Institute, which is
operated by AURA Inc., under contract from NASA.  The National Radio
Astronomy Observatory is operated by AURA Inc., under co-operative
agreement with the National Science Foundation. The United Kingdom
InfraRed Telescope is operated by the Joint Astronomy Centre on behalf of
PPARC. HJAR acknowledges support from an EU twinning project, a programme
subsidy granted by the Netherlands Organisation for Scientific Research
(NWO) and a NATO research grant.  This work was supported in part by the
Formation and Evolution of Galaxies network set up by the European
Commission under contract ERB FMRX-- CT96--086 of its TMR programme. We
thank the referee for a number of useful comments and suggestions.

\label{lastpage}
\bibliography{pnb} 
\bibliographystyle{mn} 

\end{document}